\documentclass[aps, prd, amsmath, amssymb, amsfonts, floats, %
floatfix, superscriptaddress, nofootinbib, twocolumn, showpacs]%
{revtex4}

\allowdisplaybreaks[2]

\usepackage{graphicx}
\usepackage[usenames, dvipsnames]{color}
\usepackage[colorlinks, pdfborder={0 0 0}, plainpages=false]{hyperref}
\usepackage{breakurl}
\usepackage{amsmath, amssymb, amsfonts}
\usepackage{xspace} 
\usepackage{dcolumn}
\usepackage{bm}
\usepackage{multirow} 

\def\laq{\raise 0.4ex\hbox{$<$}\kern -0.8em\lower 0.62ex\hbox{$\sim$}}
\def\gaq{\raise 0.4ex\hbox{$>$}\kern -0.7em\lower 0.62ex\hbox{$\sim$}}

\definecolor{CiteColor}{rgb}{0, 0.5, 0} %
\hypersetup{citecolor=CiteColor} %
\definecolor{RefColor}{rgb}{0.55, 0, 0} %
\hypersetup{linkcolor=RefColor} %

\usepackage{ulem}
\definecolor {darkgreen}{rgb}{0.2, 0.7, 0.2}

\newcommand{\MIT}{\affiliation{Department of Physics and MIT Kavli
    Institute, 77 Massachusetts Avenue, Cambridge, MA 02139}}
\newcommand{\Darth}{\affiliation{Department of Physics, University of
    Massachusetts Dartmouth, North Dartmouth, MA 02747}}
\newcommand{\CITA}{\affiliation{Canadian Institute for Theoretical
    Astrophysics, University of Toronto, 60 St.\ George St., Toronto,
    ON M5S 3H8, Canada}}
\newcommand{\Perim}{\affiliation{Perimeter Institute for Theoretical
    Physics, Waterloo, ON N2L 2Y5, Canada}}
\newcommand{\Maryland}{\affiliation{Maryland Center for Fundamental
    Physics \& Joint Space-Science Institute,\\ Department of Physics,
    University of Maryland, College Park, MD 20742, USA}}

\begin{document}

\title{Modeling the horizon-absorbed gravitational flux for equatorial-circular orbits in Kerr spacetime}

\author{Andrea Taracchini} \Maryland %
\author{Alessandra Buonanno} \Maryland %
\author{Scott A.\ Hughes} \MIT \CITA \Perim %
\author{Gaurav Khanna} \Darth %

\begin{abstract}
  We propose an improved analytical model for the horizon-absorbed
  gravitational-wave energy flux of a small body in circular orbit in
  the equatorial plane of a Kerr black hole. Post-Newtonian (PN)
  theory provides an analytical description of the multipolar
  components of the absorption flux through Taylor expansions in the
  orbital frequency. Building on previous work, we construct a
  mode-by-mode factorization of the absorbed flux whose Taylor
  expansion agrees with current PN results.  This factorized form
  significantly improves the agreement with numerical results obtained
  with a frequency-domain Teukolsky code, which evolves through a
  sequence of circular orbits up to the photon orbit. We perform the
  comparison between model and numerical data for dimensionless Kerr
  spins $-0.99 \leq q \leq 0.99$ and for frequencies up to the light
  ring of the Kerr black hole. Our proposed model enforces the
  presence of a zero in the flux at an orbital frequency equal to the
  frequency of the horizon, as predicted by perturbation theory. It
  also reproduces the expected divergence of the flux close to the
  light ring.  Neither of these features are captured by the
  Taylor-expanded PN flux. Our proposed absorption flux can also help
  improve models for the inspiral, merger, ringdown of small
  mass-ratio binary systems.
\end{abstract}

\date{\today}


\maketitle

\section{Introduction}
\label{sec:intro}

Extreme-mass-ratio inspirals (EMRIs) are among the most interesting
candidate sources for future space-based gravitational wave (GW)
detectors. In these systems a particle/small body, like a star or a
black hole (BH), orbits a supermassive BH and spirals in due to energy
losses in GWs. Computational modeling of EMRIs is uniquely challenging
due to the long duration and the high level of accuracy required in
the waveforms for the purposes of
detection~\cite{AmaroSeoane:2007aw}. This implies that the orbital
dynamics needs to be computed over long time intervals with sufficient
accuracy. To lowest order in the mass ratio, EMRIs can be described
using black hole perturbation theory to compute how the ``self force''
produced by the small body's interacts with its own spacetime
deformation (see, e.g., Refs.\ {\cite{Barack:2009ux, Poisson:2011nh}}
for recent reviews).  If the system evolves slowly enough, the impact
of {\em dissipative} self forces can be described using the Teukolsky
equation~\cite{Teukolsky:1973ha} to compute the slowly-changing
evolution of the integrals of Kerr geodesic orbits (i.e., an orbit's
energy, angular momentum, and Carter constant).  The inspiral is then
well described by a slowly evolving sequence of geodesic orbits.  In
Refs.~\cite{Cutler93, Cutler:1994pb, Hughes:1999bq, Hughes:2001jr,
  Glampedakis:2002ya, Drasco:2005kz}, this approach has been pursued
through purely numerical schemes.
  
Purely analytical approaches and modeling are also viable. Since the
motion of the particle eventually becomes significantly relativistic,
a post-Newtonian (PN) treatment~\cite{Sasaki:2003xr, Futamase:2007zz,
  Blanchet:2006zz} of this problem (taking the limit of small mass
ratio) is bound to fail towards the end of the inspiral. In fact, PN
theory used for long-time integration of EMRIs leads to significant
discrepancies in the number of orbital cycles. These accumulate rather
uniformly during the inspiral, even before reaching the innermost
stable circular orbit (ISCO)~\cite{Mandel:2008bc}.  More suitable
approaches are BH perturbation theory and the self-force
formalism~\cite{Teukolsky:1973ha,Poisson:2011nh}, which include all
relativistic effects but expand in the small mass-ratio parameter.

In this work we focus on a specific aspect of the problem, namely the
GW energy flux absorbed by the BH horizon. The particle orbiting the
central Kerr BH radiates GWs which partly leave the binary towards
null infinity (and constitute the so-called flux at infinity), and
partly fall into the event horizon (and constitute the so-called
absorption flux). Interest in the absorption flux was shown as early
as the 70's, when Ref.~\cite{Misner:1972kx} investigated its possible
impact on the dynamics of bodies in the vicinity of the supermassive
BH at the center of our galaxy.

For some orbits and black-hole spins, the absorption of GWs by the
event horizon can be described as a Penrose-like
process~\cite{Penrose:1969pc}, i.e., as the extraction of rotational
energy of the Kerr BH by means of negative-energy GWs.  The
``absorbed'' flux in these cases is actually negative.
Reference~\cite{Press:1972zz} formally suggested this Penrose-like
interpretation for scalar (instead of gravitational) perturbations of
a Kerr BH using the Teukolsky equation. The authors also looked for
orbits which would have a perfect balance between the energy losses in
scalar waves to infinity and the aforementioned energy extraction.
Such orbits would have a constant radius, and were named ``floating
orbits"\footnote{Similar behavior was noted by Hod in the context of
  massive-scalar fields, so-called ``stationary
  clouds''~\cite{Hod2012}.}. Subsequently,
Ref.~\cite{Teukolsky:1974yv} extended the calculation of the ingoing
energy flux to gravitational perturbations of a Kerr BH [see in
  particular Eq.~(4.44) therein], and computed it numerically for
different values of the spin of the central object [see Fig.~2 in
  Ref.~\cite{Teukolsky:1974yv}].  Reference \cite{Detweiler:1978ge}
later definitively ruled out the existence of floating orbits in the
case of gravitational perturbations.  More recent work
{\cite{Kapadia:2013kf}} suggests that floating orbits can only exist
around central bodies with an extremely unusual multipolar structure.

Further insight into the horizon-absorbed flux in a BH binary system
can be gained from a parallel with the phenomenon of tides. In the
early 70's, Refs.~\cite{Hartle:1973zz,Hartle:1974gy} computed how a
stationary particle tidally perturbs a slowly rotating Kerr BH,
finding that the BH dissipates energy by spinning down.  The same
phenomenon happens in a Newtonian binary system, such as when a moon
perturbs a slowly rotating planet (treated as a fluid body with
viscosity).  This phenomenon is known as ``tidal heating.''  Somewhat
remarkably, there is a close analogy between the spindown of a black
hole and the spindown of a fluid body due to the tidal interaction:
The tidal interaction raises a bulge on the black hole's event
horizon, and one can regard that bulge as exerting a torque on the
orbit.  This torque spins up or spins down the hole, depending on the
relative frequency of the orbit and the hole's rotation.  Using the
membrane paradigm {\cite{Thorne-Price-MacDonald}}, one can even
associate an effective viscosity to the black hole.  The hole's
viscosity relates the rate at which the horizon's generators are
sheared to the rate at which the hole's area (or entropy) is
increased.  The black hole's viscosity plays an important role in
determining the geometry of the hole's bulge, much as the viscosity of
a fluid body in part determines the geometry of its tidal bulge.

A renewed interest in the BH-absorption flux was rekindled in the
90's, when, using BH perturbation theory, Ref.~\cite{Poisson:1994yf}
computed in full analytical form the leading-order absorption flux for
a particle in circular orbit around a Schwarzschild BH.  These initial
results indicated that the horizon flux is suppressed relative to the
flux to infinity by a factor of $v^8$, where $v$ is the orbital speed.
This result was then generalized to the spinning case in
Refs.~\cite{Tagoshi:1997jy, Mino:1997bx}, where the ingoing flux was
computed up to 6.5PN order beyond the leading order luminosity at
infinity.  Spin dramatically changes the leading impact of the horizon
flux: The suppression factor becomes $(v^3 - q)v^5$ (where $q \equiv
a/M$ is the Kerr parameter per unit mass).  Numerical studies of
strong field radiation reaction showed that neglect of the horizon
flux would introduce large errors into Kerr inspiral models --- many
thousands of radians for inspiral into rapidly rotating black holes
{\cite{Hughes:2001jr}}.

The extension to comparable-mass BH binaries was first attempted in 
Ref.~\cite{Alvi:2001mx}, which computed the changes in mass and
angular momentum of the holes up to 4PN order beyond the leading order 
luminosity at infinity. Reference~\cite{Poisson:2004cw}
constructed a general approach to this problem, deriving formulae for
the flow of energy and angular momentum into a BH as functions of the
generic tidal fields perturbing it. This formalism was applied in
Ref.~\cite{Taylor:2008xy} to the specific tidal environment of a
comparable-mass binary in the slow-motion approximation, allowing the computation of the spinning
absorption fluxes to higher PN order than Ref.~\cite{Alvi:2001mx}.
Recently Ref.~\cite{Chatziioannou:2012gq} pushed the calculation of Ref.~\cite{Taylor:2008xy} to an even higher PN order.

In recent years, significant effort has been put into improving the
analytical modeling of the GW fluxes, both ingoing and at infinity,
with respect to the exact, numerical solution of the Teukolsky
equation. In particular, Refs.~\cite{Damour2007,DIN} proposed a
factorization of the Taylor-expanded PN formulae for the flux at infinity
in the Schwarzschild case, improving the agreement with the numerical
data. Reference~\cite{Pan2010hz} extended this approach to the spinning
case. Later on Ref.~\cite{Nagar:2011aa}
applied the same idea of factorizing the PN Taylor-expanded PN predictions
to the absorption flux in the nonspinning limit, extending the model
also to comparable-mass binaries. Our work has the primary goal of
studying the factorization of the BH-absorption flux for the Kerr
case. The orbits we consider are circular and lie
in the equatorial plane of the central, rotating BH. The 
PN-expanded formulae for the spinning absorption flux can be found in
Refs.~\cite{Tagoshi:1997jy, Mino:1997bx}.

An improved analytical modeling of the GW fluxes in the test-particle
limit is crucial because of the practical need for fast generation
of reliable time-domain waveforms for these 
systems. Several papers~\cite{Bernuzzi:2010ty,Bernuzzi:2010xj,Yunes:2010zj,
Bernuzzi:2011aj,Barausse:2011kb,Bernuzzi:2012ku} have 
already incorporated analytical fluxes into effective-one-body (EOB)
models for EMRIs. One solves the Hamilton equations for the Kerr Hamiltonian with dissipation 
effects introduced through a radiation-reaction force that is proportional 
to the GW flux. As far as the ingoing flux is concerned, Ref.~\cite{Yunes:2010zj} worked with
spinning EMRIs, including the BH-absorption terms in Taylor-expanded PN 
form~\cite{Tagoshi:1997jy,Mino:1997bx}. The authors of Ref.~\cite{Bernuzzi:2012ku} focussed on the
nonspinning case, and used the factorized nonspinning absorption flux
of Ref.~\cite{Nagar:2011aa}. Our work can be regarded as a step
beyond Ref.\ \cite{Yunes:2010zj} toward building a high-quality EOB
model for EMRIs with spinning black holes.  Besides the specific
problem of the long inspiral in EMRIs, the EOB model has proven effective in
describing the whole process of inspiral, merger and ringdown --- 
for example Ref.~\cite{TaracchiniSpinTPL} has used the results of this work to 
model merger waveforms from small mass-ratio binary systems for any BH spin.  

This paper is organized as follows. In Sec.~\ref{sec:NumFlux} we
discuss the numerical computation of energy fluxes at infinity and
into the BH horizon using the frequency-domain Teukolsky equation.  We
investigate the behavior of these fluxes close to the photon orbit,
discussing their main features. In Sec.~\ref{sec:Fact} we review the
factorization of the analytical GW fluxes computed in PN theory and
apply it to the spinning BH-absorption flux. In
Sec.~\ref{sec:Comparisons} we show comparisons of the factorized and
Taylor-expanded PN fluxes to the numerical fluxes. In
Sec.~\ref{sec:Conclusions} we conclude and discuss future research.
Appendix~\ref{app:LRsource} discusses in more depth aspects of the
near-light-ring fluxes, in particular how these fluxes diverge at the
photon orbit, and how this divergence can be analytically factored
from the fluxes. Appendix~\ref{sect:Appendixflm} contains the explicit
formulae for a particular choice of the factorization model of the
BH-absorption flux. Lastly, in Appendices~\ref{sect:AppendixFitFInf}
and \ref{sect:AppendixFitFH} we provide fits to the Teukolsky-equation
fluxes that can be employed for accurate evolution of EMRIs or
inspiral, merger and ringdown waveforms for small mass-ratio binary
systems.

Throughout this paper, we use geometrized units with $G = c = 1$.  We
use $\mu$ to label the mass of the small body; $M$ and $q \equiv a/M$
are the mass and dimensionless spin of the Kerr black hole,
respectively.  The spin parameter $q$ ranges from $-1$ to $+1$, with
positive values describing prograde orbits, and negative values
retrograde ones.  With this convention, the orbital angular momentum
$L_z$ and orbital frequency $\Omega$ are always positive.  When we
discuss radiation and fluxes, we will often decompose it into modes.
Through most of the paper, we decompose the radiation using spheroidal
harmonics $S_{\ell m\omega}(\theta,\phi)$, discussed in more detail in
Sec.\ \ref{sec:NumFlux}.  In Sec.\ {\ref{sec:Fact}}, we will find it
useful to use an alternative decomposition into spherical harmonics,
$Y_{lm}(\theta,\phi)$.  We will strictly use the harmonic indices
$(\ell, m)$ for spheroidal harmonics, and $(l,m)$ for spherical
harmonics.

\section{Numerical computation of the gravitational-wave fluxes}
\label{sec:NumFlux}

In this section we first outline how we numerically compute GW
fluxes (both ingoing and at infinity) by solving the frequency-domain
Teukolsky equation.  Much of this has been described in detail in
other papers, in particular,
Refs.~\cite{Hughes:1999bq,Drasco:2005kz}, so our discussion just
highlights aspects which are crucial to this paper.  Then, we discuss
the main characteristics of those fluxes, their strength as function
of the spin and their behavior close to the photon orbit.
 
\subsection{Synopsis of numerical method}
\label{sec:numericalsynopsis}

The Teukolsky ``master'' equation is a partial differential equation
in Boyer-Lindquist coordinates $r$, $\theta$, and $t$ (the axial
dependence is trivially separated as $e^{im\phi}$).  It describes the
evolution of perturbing fields of spin weight $s$ to a Kerr black hole
{\cite{Teukolsky:1973ha}}.  The equation for $s = -2$ describes the
curvature perturbation $\psi_4$, a projection of the Weyl curvature
tensor which represents outgoing radiation.  With some manipulation,
solutions for $s = -2$ give radiation at the hole's event horizon as
well {\cite{Teukolsky:1974yv}}.

The master equation for $s = -2$ separates by introducing the
multipolar decomposition
\begin{eqnarray}
\psi_4 &=& \frac{1}{(r - iMq\cos\theta)^4}\int_{-\infty}^\infty
d\omega
\nonumber\\
& &\times \sum_{\ell m} R_{\ell m\omega}(r)
S^-_{\ell m\omega}(\theta,\phi)e^{-i\omega t}\;.
\label{eq:psi4decomp}
\end{eqnarray}
Here and elsewhere in this paper, any sum over $\ell$ and $m$ is
taken to run over $2 \le \ell < \infty$, and $-\ell \le m \le \ell$,
unless explicitly indicated otherwise.  The function $S^-_{\ell
  m\omega}(\theta,\phi)$ is a spheroidal harmonic of spin-weight $-2$;
the minus superscript is a reminder of this spin weight.  It reduces
to the spin-weighted spherical harmonic when $qM\omega = 0$:
$S^-_{\ell m\omega}(\theta,\phi) = Y^-_{\ell m}(\theta,\phi)$ in this
limit.  The radial dependence $R_{\ell m\omega}(r)$ is
governed by the equation
\begin{equation}
\Delta^2\frac{d}{dr}\left(\frac{1}{\Delta}\frac{dR_{\ell m\omega}}{dr}\right)
-V(r)R_{\ell m\omega} = -{\cal T}_{\ell m\omega}(r)\;.
\label{eq:FDteuk}
\end{equation}
The quantity $\Delta = r^2 - 2Mr + M^2q^2$, and the potential $V(r)$
can be found in Refs.\ {\cite{Hughes:1999bq,Drasco:2005kz}}.  Note
that in Eqs.\ (\ref{eq:psi4decomp}), (\ref{eq:FDteuk}),
(\ref{eq:pointsource}), and (\ref{eq:particularsolution}), the
variable $r$ labels the coordinate of an arbitrary field point.  This
is true only in these specific equations; elsewhere in this paper, $r$
gives the radius of a circular orbit.

Equation (\ref{eq:FDteuk}) is often called the frequency-domain
Teukolsky equation, or just the Teukolsky equation.  The source ${\cal
  T}_{\ell m\omega}(r)$ is built from certain projections of the
stress-energy tensor for a small body orbiting the black hole:
\begin{equation}
T_{\alpha\beta} = \frac{\mu u_\alpha u_\beta}{\Sigma\sin\theta
  (dt/d\tau)} \delta[r - r_{\rm o}(t)] \delta[\theta - \theta_{\rm
    o}(t)] \delta[\phi - \phi_{\rm o}(t)]\;.
\label{eq:pointsource}
\end{equation}
The subscript ``o'' means ``orbit,'' and labels the coordinates of an
orbiting body's worldline.  We focus on circular equatorial orbits, so
$\theta_{\rm o}(t) = \pi/2$, and $r_{\rm o}(t) = \mbox{constant}$.
Notice the factor $(dt/d\tau)^{-1}$ that appears here.  As the light
ring (LR) is approached, $dt/d\tau \to 0$, and this factor introduces
a pole into the energy fluxes.  We discuss the importance of this pole
in more detail below, and describe how it can be analytically factored
from the fluxes in Appendix {\ref{app:LRsource}}.

We consider orbits from $r_{\rm o}$ near the light ring out to very
large radius ($r_{\rm o} \simeq 10^4M$). Previous work has typically
only considered orbits down to the ISCO.  However, our code can solve
Eq.\ (\ref{eq:FDteuk}) for any bound orbit, including unstable
ones\footnote{In Ref.\ {\cite{Yunes:2010zj}}, we stated that our code
  did not work inside the ISCO because there are no stable orbits
  there.  It is true that we cannot relate the fluxes to quantities
  like the rate of change of orbital radius, inside the ISCO, but the
  code can compute fluxes from unstable orbits perfectly well in this
  regime.}.  No modifications are needed to broaden our study to these
extremely strong-field cases, though there are some important
considerations regarding convergence, which we discuss below. 

We solve Eq.\ (\ref{eq:FDteuk}) by building a Green's function from
solutions to the homogeneous equation (i.e., with ${\cal T}_{\ell
  m\omega} = 0$) and then integrating over the source; see
Refs.\ \cite{Hughes:1999bq,Drasco:2005kz} for details.  The resulting
solutions have the form
\begin{equation}
R_{\ell m\omega}(r) = \left\{ \begin{array}{ll}
Z^{\rm H}_{\ell m\omega} R^\infty_{\ell m\omega}(r)
& \mbox{\ \ \ \  $ r \to \infty$,} \\
\\
Z^\infty_{\ell m\omega} R^{\rm H}_{\ell m\omega}(r)
& \mbox{\ \ \ \ $r \to r_+$}, \\
\end{array}
\right.
\label{eq:particularsolution}
\end{equation}
where
\begin{eqnarray}
Z^{\rm H}_{\ell m\omega} &=& C^{\rm H} \int_{r_+}^{r_{\rm orb}}
dr'\frac{R^{\rm H}_{\ell m\omega}(r'){\cal
    T}_{\ell m\omega}(r')}{\Delta(r')^2}\;,
\label{eq:Z1H}\\
Z^\infty_{\ell m\omega} &=& C^\infty \int_{r_{\rm orb}}^\infty
dr'\frac{R^\infty_{\ell m\omega}(r'){\cal
    T}_{\ell m\omega}(r')}{\Delta(r')^2}\;,
\label{eq:Z1Inf}
\end{eqnarray}
and where $R^\star_{\ell m\omega}(r)$ are the homogeneous solutions
from which we build the Green's function ($\star$ means $\infty$ or
$H$, as appropriate).  The symbol $C^\star$ is shorthand for a
collection of constants whose detailed form is not needed here (see
Sec.\ III of Ref.~{\cite{Drasco:2005kz}} for further discussion).

The code we use to compute these quantities is described in
Refs.\ {\cite{Hughes:1999bq, Drasco:2005kz}}, updated to use the
methods introduced by Fujita and Tagoshi~\cite{Fujita:2004rb,
  Fujita:2009uz} (see also Ref.\ \cite{Sasaki:2003xr}).  This method
expands the homogeneous Teukolsky solutions as a series of
hypergeometric functions, with the coefficients of these series
determined by a three term recurrence relation, Eq.\ (123) of
Ref.\ {\cite{Sasaki:2003xr}}.  Successfully finding these coefficients
requires that we first compute a number $\nu$ which determines the
root of a continued fraction equation, Eq.\ (2.16) of
Ref.\ {\cite{Fujita:2004rb}}.  Provided we can find $\nu$, we
generally find very accurate\footnote{We estimate our solutions to
  have a fractional error $\sim 10^{-14}$ in these cases.  R.\ Fujita
  has provided numerical data computed with an independent Teukolsky
  solver.  We find 15 or more digits of agreement in our computed
  amplitudes in all cases.} solutions for $R^\star_{lm\omega}$.
However, there are some cases where we cannot compute $\nu$, typically
close to the light ring for $\ell \gtrsim 60$ (although these
difficulties arise at smaller $\ell$ for large spin, retrograde orbits
near the light ring).  In these cases, the root of the continued
fraction lies very close to a pole of this equation.  (Figures 4 and 5
of Ref.\ {\cite{Fujita:2004rb}} show examples of the pole and root
structure of this equation for less problematic cases.)  We discuss
where this limitation impacts our analysis below. 

For periodic orbits, the coefficients $Z^\star_{\ell m\omega}$ have
a discrete spectrum:
\begin{equation}
Z^\star_{\ell m\omega} = Z^\star_{\ell m}\delta(\omega - \omega_m)\;,
\end{equation}
where $\omega_m = m\Omega$, with $\Omega$ the orbital frequency of the
small body.  The amplitudes $Z^\star_{\ell m}$ then completely
determine the fluxes of energy and angular momentum:
\begin{eqnarray}
\dot E^\infty &=& \sum_{\ell m} \frac{|Z^{\rm H}_{\ell m}|^2}{4\pi\omega_m^2}
\nonumber\\
&\equiv& \sum_{\ell m}F^\infty_{\ell m,\rm Teuk} = F^\infty_{\rm Teuk}\;,
\label{eq:EdotInf_spheroid}
\\
\dot E^{\rm H} &=& \sum_{\ell m}
\frac{\alpha_{\ell m}|Z^\infty_{\ell m}|^2}{4\pi\omega_m^2}
\nonumber\\
&\equiv& \sum_{\ell m}F^{\rm H}_{\ell m,\rm Teuk} = F^{\rm H}_{\rm Teuk}\;. 
\label{eq:EdotH_spheroid}
\end{eqnarray}
For circular and equatorial orbits, fluxes of angular momentum are
simply related to energy fluxes: $\dot E^\star = \Omega \dot
L^\star$.

The factor $\alpha_{\ell m}$ which appears in fluxes on the horizon
arises from converting the curvature scalar $\psi_4$ to $\psi_0$ in
order to determine, via the area theorem, the rate at which the black
hole's mass and spin change due to tidal coupling with the orbiting
body (see Ref.\ \cite{Teukolsky:1974yv} for discussion). The fluxes
carried by radiation are then determined by imposing global
conservation of energy and angular momentum\footnote{Our ability to
  use these conservation laws follows from the fact that the Kerr
  spacetime admits timelike and axial Killing vectors.}.  This factor
is given by
\begin{equation}\label{alphalm}
\alpha_{\ell m} = \frac{256(2Mr_+)^5p_m(p_m^2 + 4\epsilon^2)(p_m^2 +
  16\epsilon^2)\omega_m^3}{|c_{\ell m}|^2}\;,
\end{equation}
where $r_{+}/M=1+\sqrt{1-q^{2}}$ and $M\Omega_{\rm H}=q/(2r_{+})$ are the radial position and frequency of the event horizon, $p_m = \omega_m - m\Omega_{\rm H}$, $\epsilon = \sqrt{1 -
  q^2}/(4r_+)$, and
\begin{eqnarray}
|c_{\ell m}|^2 &=& \left[(\lambda+2)^2 + 4qM\omega_m -
  4q^2M^2\omega_m^2\right]
\nonumber\\
&\times& (\lambda^2 + 36mqM\omega_m - 36q^2M^2\omega_m^2)
\nonumber\\
&+& (2\lambda+3) (96q^2M^2\omega_m^2 - 48mqM\omega_m)
\nonumber\\
&+& 144M^2\omega_m^2(1 - q^2)\;.
\label{eq:Clmdef}
\end{eqnarray}
In this quantity,
\begin{equation}
\lambda = {\cal E}_{\ell m} - 2qMm\omega_m + q^2M^2\omega_m^2 -2\;.
\end{equation}
(Note that the subscript was incorrectly left off of $\omega_m$ when
$\lambda$ was defined in Ref.\ {\cite{Yunes:2010zj}}.)  The number
${\cal E}_{\ell m}$ is the eigenvalue of the spheroidal harmonic; in the
Schwarzschild limit, it reduces to $\ell(\ell + 1)$.  Notice that
$\alpha_{\ell m} \propto p_m \propto (\Omega - \Omega_H)$.  This means
that the horizon flux is negative when $\Omega < \Omega_H$, consistent
with the leading order result, Eq.\ (\ref{FHLO}).

All the data computed with these methods will be referred to as
``numerical data'' in the rest of the paper.

\subsection{Discretization of orbits and convergence of the flux sums}
\label{sect:TeukSums}
We compute these fluxes on a pair of grids evenly spaced in the
velocity variable
\begin{equation}\label{KeplerKerr}
v \equiv (M\Omega)^{1/3} = \left[(r/M)^{3/2} + q\right]^{-1/3}\;.
\end{equation}
(In this section and beyond, there is no longer an ambiguity between
labels for field point or orbital radius.  In the remainder of the
paper, $r$ will label the radius of a circular orbit.)  Our ``outer''
grid consists of $10^4$ points spaced from $v = 0.01$ ($r \simeq
10^4M$) to the ISCO radius~{\cite{Bardeen:1972fi}},
\begin{eqnarray}
\frac{r_{\rm ISCO}}{M} &=& 3 + Z_2 \mp \sqrt{(3 - Z_1)(3 + Z_1 + 2Z_2)}\;,
\nonumber\\
Z_1 &=& 1 + (1 - q^2)^{1/3}\left[(1 + q)^{1/3} + (1 - q)^{1/3}\right]\;,
\nonumber\\
Z_2 &=& (3q^2 + Z_1^2)^{1/2}\;.
\label{eq:iscorad}
\end{eqnarray}
[The upper sign in Eq.\ (\ref{eq:iscorad}) is for prograde orbits, $q
  > 0$, and the lower for retrograde, $q < 0$.]  Our ``inner'' grid
consists of 100 points spaced from the ISCO to just outside the light
ring: $r_{\rm min} = r_{\rm LR} + 0.01M$, where
{\cite{Bardeen:1972fi}}
\begin{equation}
\frac{r_{\rm LR}}{M} = 2\left [1 + \cos\left( \frac{2}{3}\arccos(-q)\right)\right ]\;.
\label{eq:lightring}
\end{equation}
In some cases, we put $r_{\rm min} = r_{\rm LR} + 0.009M$.  This is to
avoid the problem mentioned in the text following Eq.\ (\ref{eq:Z1Inf}):
For very strong field (large $\Omega$) orbits, when $\ell \gtrsim 60$,
we sometimes find a value of $m\Omega$ for which we cannot find the
number $\nu$, and hence cannot solve the Teukolsky equation.  We find
empirically that modifying the grid slightly to avoid those
problematic frequencies fixes this problem in many cases.

For circular, equatorial orbits, the largest contributions to the sums
for $F^\star$ tend to come at small $\ell$ (usually $\ell = 2$), and
then fall off as explained in Eq.~(\ref{Fell}) as we go to higher
values of $\ell$.  We consider a sum to have ``converged'' when we
reach a value $\ell \equiv \ell_{\rm max}$ such that the fractional
change in the sum due to all terms with $\ell = \ell_{\rm max}$ is
smaller than $10^{-14}$ for three consecutive values of $\ell$.  This
criterion was also used in Ref.\ {\cite{Yunes:2010zj}}.  For all
orbits up to and including the ISCO, we were able to achieve this
convergence for every spin that we examined.  However, the $\ell_{\rm
  max}$ needed varies considerably with spin, mostly because the
location of the ISCO varies strongly with spin: The deeper into the
strong field we must go, the more multipoles are needed for
convergence.  For Schwarzschild, convergence required going to
$\ell_{\rm max} = 30$ at the ISCO.  For prograde $q = 0.99$, the same
level of convergence took us to $\ell_{\rm max} = 66$ at the ISCO.

We were unable to achieve this convergence criterion for all orbits
inside the ISCO.  As we approach the light ring, the falloff of
contributions to the flux sums becomes shallow, and the number of
multipoles needed to converge becomes extremely large.  At our
innermost gridpoint $r_{\rm min}$, for $\ell\sim 70$ we find
\begin{equation}
\label{FellRatio}
\frac{F^\star_{\ell}}{F^\star_{\ell - 1}} \simeq 1 - \epsilon\;,
\end{equation}
where $F^\star_{\ell}\equiv\sum_{m}{F^\star_{\ell m}}$, $\epsilon
\approx \mbox{a few}\times 0.01$.  This is consistent with past
analytical work on geodesic synchrotron radiation~\cite{Davis:1972dm,
  Breuer:1973kt, Chrzanowski:1974nr,BreuerBook} which showed that a
similar flux quantity (defined by summing over all allowed values of
$\ell$ for a fixed $m$) is proportional to $(m_{\rm
  c}/m)\exp{(-2m/m_{\rm c})}$, where
\begin{equation}
m_{\rm c} \equiv \frac{2\sqrt{3}}{\pi} \frac{r_{\rm LR}/M +3}{\sqrt{r_{\rm LR}/M}} \left(\frac{E}{\mu}\right)^{2}\,,
\end{equation}
and $E$ is the binding energy for circular orbits given in
Eq.~(\ref{Ecirc}), which diverges at the light ring as $(r - r_{\rm LR})^{-1/2}$.  The sums are
dominated by the $\ell = |m|$ contributions, so either limiting form
--- $(m_{\rm c}/m) \exp{(-2m/m_{\rm c})}$ or $(\ell_{\rm c}/\ell)
\exp{(-2\ell/\ell_{\rm c})}$ --- is accurate.  In our case, we find
\begin{equation}\label{Fell}
F^{\infty}_{\ell} \propto \frac{(E/\mu)^{2}}{\ell} \exp{\left[-2\ell \left(\frac{r}{r_{\rm LR}}-1\right)\right]}\,,
\end{equation}
where $E$ is the energy of the circular orbit at radius $r$, given by
Eq.\ (\ref{Ecirc}) below. It was shown that the same result holds also
for the absorption flux for orbits close to the photon orbit.  When
$r=r_{\rm min}$ the exponential factor is $\approx 1$ up to $\ell \sim
\mathcal{O}((r_{\rm min}-r_{\rm LR})^{-1})\gtrsim 100 $, which is
consistent with the behavior described by Eq.\ (\ref{FellRatio}).
These flux sums would converge eventually if we computed enough
multipolar contributions.  However, at very large values of $\ell$ and
$m$, the methods we use to solve for the homogeneous Teukolsky
solutions $R^\star_{\ell m\omega}(r)$ fail to find a solution.  For
all prograde orbits, we terminate the flux sums at $\ell = 70$ if the
convergence criterion has not been met at this point.  Large $q$
retrograde orbits are more of a challenge; we have difficulty
computing these modes (for the reasons discussed in
Sec.\ {\ref{sec:numericalsynopsis}} above) for somewhat smaller values
of $\ell$ for large, negative $q$.  We terminate our sums when we
cannot reliably compute $R^\star_{\ell m\omega}(r)$.  The value of
$\ell$ we reach is shown in Table {\ref{tab:lightringconv}}, and
varies from $70$ for $q = -0.5$ to $43$ for $q = -0.99$.

\begin{table*}
\begin{ruledtabular}
\begin{tabular}{c|ccccc}
$q$ & $\ell_{\rm max}$ & $F^\infty_{\ell = \ell_{\rm max}}/F^\infty$ & $F^{\rm
    H}_{\ell = \ell_{\rm max}}/F^{\rm H}$ &
  $\varepsilon^\infty_{\rm negl}$ & $\varepsilon^{\rm H}_{\rm
    negl}$ \\[2pt]
\hline\\[-8pt]%
$0 .99 $ & $70$ & $7 .06\times 10^{-5} $ & $6 .78\times 10^{-9} $ & $0.0398\%$ & $3.82 \times 10^{-6}\%$ \\[2 pt]
$0.9$ & $70$ & $6.93 \times 10^{-4}$ & $2.28 \times 10^{-4}$ & $1.10\%$ &  $0.36\%$ \\[2pt]
$0.7$ & $70$ & $1.38 \times 10^{-3}$ & $1.17 \times 10^{-3}$ & $3.54\%$ & $3.00\%$ \\[2pt]
$0.5$ & $70$ & $1.49 \times 10^{-3}$ & $1.44 \times 10^{-3}$ &$ 4.80\%$&  $4.64\%$\\[2pt]
$0.0$ & $70$ & $1.82 \times 10^{-3}$ & $2.04 \times 10^{-3}$ & $8.07\%$ & $9.05\%$\\[2pt]
$-0.5$ & $70$ & $2.03 \times 10^{-3}$ & $2.36 \times 10^{-3}$ & $10.9\%$&$12.7\%$\\[2pt]
$-0.7$ & $66$ & $2.31 \times 10^{-3}$ & $2.71 \times 10^{-3}$  & $13.1\%$ & $15.4\%$ \\[2pt]
$-0.9$ & $56$ & $3.10 \times 10^{-3}$ & $3.68 \times 10^{-3}$ & $18.1\%$&$21.5\%$\\[2pt]
$-0.99$ & $43$ & $4.75 \times 10^{-3}$ & $5.66 \times 10^{-3}$ & $23.5\%$& $28.1\%$ \\
\end{tabular}
\end{ruledtabular}
\caption{Diagnostics of convergence at our innermost gridpoint, $r_{\rm min} =
  r_{\rm LR} + 0.01M$, where the convergence is poorest. The second column lists the $\ell_{\rm max}$ where we end the sums for the total fluxes $F^{\star}$. The third column
  shows the flux to infinity in all $\ell = \ell_{\rm max}$ modes, normalized to
  the total flux (all modes up to and including $\ell = \ell_{\rm max}$).  The
  third column is the same data for the horizon flux. The fourth and fifth columns give
  the error measure $\varepsilon^\star_{\rm negl}$, defined by
  Eq.\ (\ref{eq:errorestimate}).  Convergence rapidly improves as we
  move away from this radius, with errors falling to $10^{-14}$ at radii
  $\mbox{a few} \times 0.1M$ from the light ring.}
\label{tab:lightringconv}
\end{table*}

To understand how much error we incur by terminating these sums, we
examine how the flux behaves at the innermost grid point at $\ell_{\rm
  max}$ and $\ell_{\rm max} - 1$.  The fractional error due to the
multipoles which have been neglected in our sum is 
\begin{equation}
\varepsilon^\star_{\rm negl} \equiv \frac{1}{F^\star}
\sum_{\ell = \ell_{\rm max}+1}^\infty F^\star_{\ell}\;.
\end{equation}
If we assume that $F^{\star}_\ell$ falls off as suggested by
Eq.~(\ref{Fell}) for $\ell \gtrsim \ell_{\rm max}$, this error can be
estimated to be
\begin{eqnarray}
\varepsilon^\star_{\rm negl} &=&\frac{F^{\star}_{\ell_{\rm max}}}{F^{\star}} \left[\frac{F^{\star}_{\ell_{\rm max}+1}}{F^{\star}_{\ell_{\rm max}}}+\frac{F^{\star}_{\ell_{\rm max}+2}}{F^{\star}_{\ell_{\rm max}}}+\cdots\right]\nonumber\\
&\leq& \frac{F^\star_{\ell_{\rm max}}}
           {F^\star} \sum_{\ell = \ell_{\rm max} + 1}^\infty
           \left(\frac{F^\star_{\ell}}{F^\star_{\ell- 1}}\right)^{\ell - \ell_{\rm max}}
\nonumber\\
&=&  \frac{F^\star_{\ell_{\rm max}}}
           {F^\star} \sum_{\ell = \ell_{\rm max} + 1}^\infty
           \left(\frac{\ell-1}{\ell}e^{-2/\ell_{\rm c}}\right)^{\ell - \ell_{\rm max}}\,.
\label{eq:errorestimate}
\end{eqnarray}
Equation (\ref{eq:errorestimate}) is
quite simple to compute, and is accurate enough for our purposes.

Table {\ref{tab:lightringconv}} summarizes how the fluxes behave at
our innermost data point for all the spins we have examined.  We see
that $\varepsilon_{\rm negl}$ varies from less than a percent to about
$20$--$30\%$ at the innermost grid point in our study.  The largest
errors are for the high spin retrograde cases, where we are forced to
terminate the sum relatively early.

These errors improve very rapidly as we move away from the light ring.
For the case of $q = -0.99$ (the case with the largest errors due to
neglected modes in our study), the contribution at $r \simeq r_{\rm
  LR} + 0.05M$ has $F^\infty_{\ell_{\rm max}}/F^\infty \simeq
1.16\times10^{-3}$, and $F^\infty_{\ell_{\rm max}}/F^\infty_{\ell_{\rm
    max} - 1} \simeq 0.930$; similar values describe the horizon flux
at this location.  Our rough estimate of the error falls to about
$1.5\%$, an order of magnitude smaller than at our innermost grid
point.  We typically find that neglected terms in the sum contribute
less than $10^{-14}$ to the total by the time we are $\mbox{a few}
\times 0.1M$ out from the light ring.

As was mentioned in the text following Eq.\ (\ref{eq:pointsource}),
the factor of $(dt/d\tau)^{-1}$ in the point-particle stress energy
tensor introduces a pole in the fluxes, leading to strong divergence
as a power of $1/(v - v_{\rm LR})$ as we approach the light ring.  We
have confirmed this behavior on a mode-by-mode basis, and have studied
it using a modified version of our code in which this behavior is
analytically factored from the fluxes (see Appendix
{\ref{app:LRsource}}).  Our numerical data up to $r_{\rm min}$ are
consistent with a divergence of the total fluxes of the form $\sim
(E/\mu)^{2}$.

It is worth emphasizing that if we use the WKB
approximation~\cite{Davis:1972dm,Breuer:1973kt,
  Chrzanowski:1974nr,BreuerBook} and normalize the fluxes (at infinity
or through the BH horizon) to the specific energy and compute them
exactly at the LR, we have $\left [F^\star_\ell/(E/\mu)^2
  \right]_{r_{\rm LR}} \sim 1/\ell$. Thus, in the WKB approximation
the total normalized fluxes diverge logarithmically when computed at
the LR.

\begin{figure*}[!ht]
  \begin{center}
    \includegraphics*[width=0.45\textwidth]{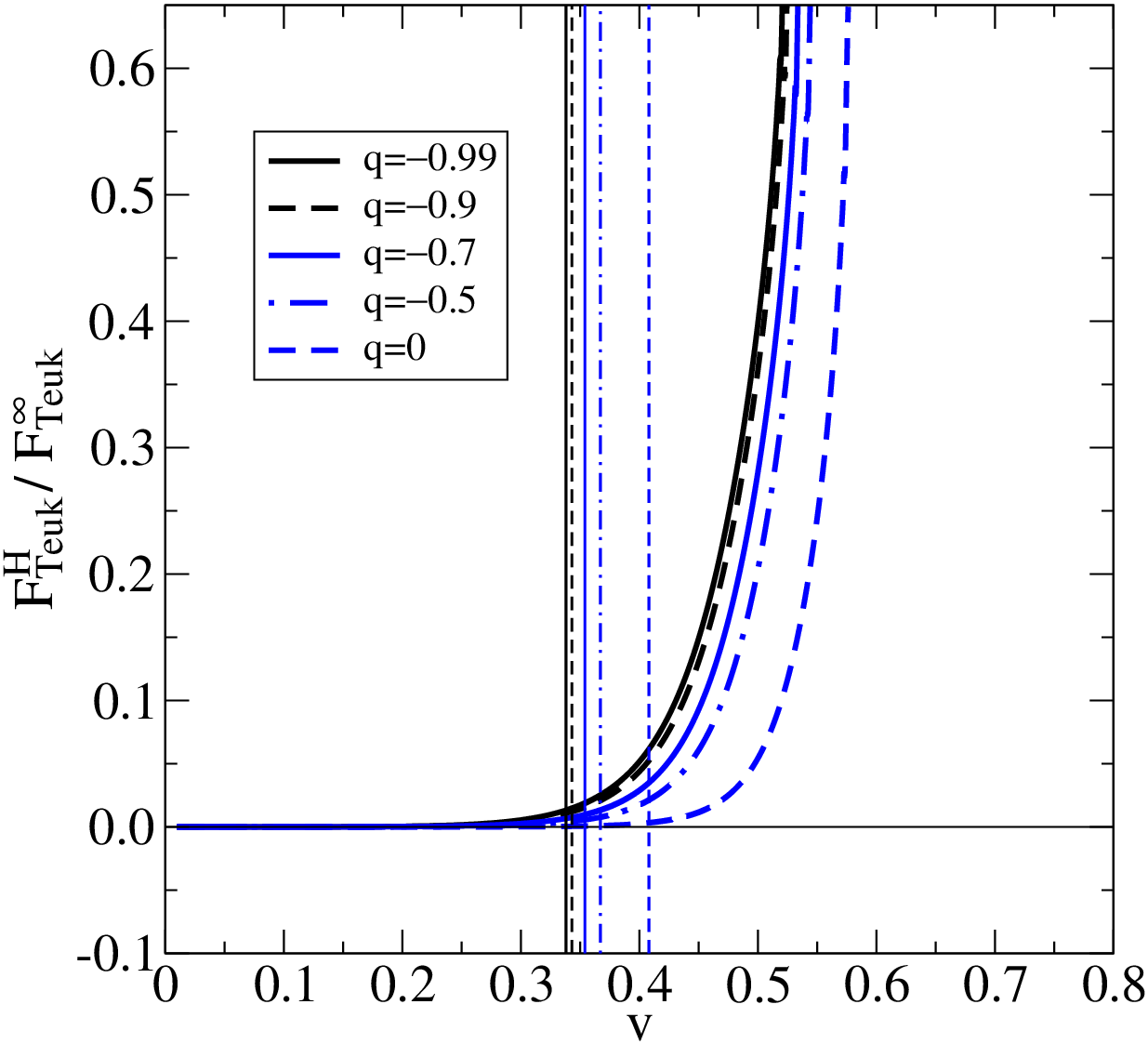}
    \hspace*{2em}
    \includegraphics*[width=0.45\textwidth]{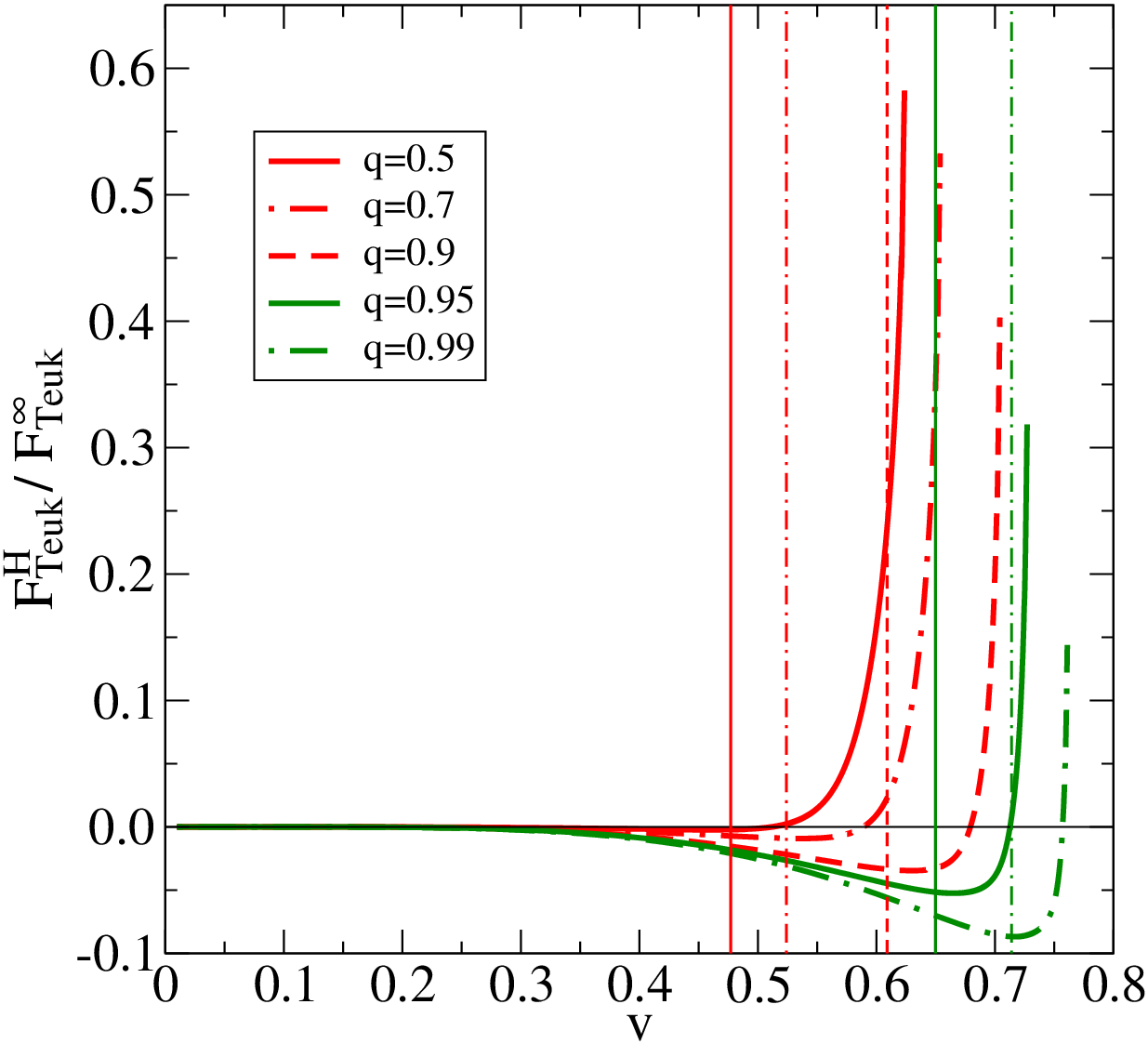}
   \caption{\label{fig:FHoverFInf}We show the ratio between the energy flux absorbed by the horizon $F^{\rm H}$ and the energy flux radiated to infinity $F^{\infty}$ for different possible values of the spin $q$, as a
    function of $v\equiv (M \Omega)^{1/3}$. The data come from the numerical solution of the 
    Teukolsky equation in the adiabatic approximation. All plots extend up to $r=r_{\rm LR}+0.01M$. 
Vertical lines mark the positions of the respective ISCOs.}
  \end{center}
\end{figure*}

\subsection{Features of numerical fluxes}
\label{subsec:features}

We now analyze the numerical fluxes and describe their main
features to gain insight for the analytical modeling. 

In Boyer-Lyndquist coordinates, at leading order  in the PN 
expansion or Newtonian order, the ingoing GW flux reads [see, e.g., Eq.~(11) in Ref.~\cite{Alvi:2001mx}]
\begin{equation}
F^{\rm H,\, N} = \frac{32}{5} \frac{\mu^2 M^6}{r^6} \Omega \left(\Omega - \Omega_{\rm H}\right)\,,\label{FHLO}
\end{equation}
where $r$ is the radial separation and $\Omega$ is the orbital frequency
of the particle. This can be compared to the leading-order  
luminosity at infinity in GWs \cite{Blanchet:2006zz}
\begin{equation}
F^{\infty,\, \rm N} = \frac{32}{5}\mu^2 r^{4}\Omega^{6}\,.\label{FInfLO}
\end{equation}
For quasi-circular inspiral, Eqs.\ (\ref{FHLO}) and (\ref{FInfLO})
tell us that $F^{\rm H,\,N}/F^{\infty,\, \rm N} \sim (M\Omega)^{5/3}$
for $q \ne 0$, so the horizon flux is 2.5PN orders beyond the flux to
infinity.  In the nonspinning limit, $F^{\rm H,\, N}/F^{\infty,\, \rm
  N} \sim (M \Omega)^{8/3}$ --- 4PN order in this case. Note that to
obtain these ratios we used Eq.~(\ref{KeplerKerr}).

Thus, at leading order the absorption flux is suppressed with respect
to the flux at infinity by $\mathcal{O}((M\Omega)^{5/3})$ for $q\neq0$
or by $\mathcal{O}((M\Omega)^{8/3})$ for $q=0$. In order to have a
more accurate assessment of the relative importance of $F^{\rm H}$ and
$F^{\infty}$, in Fig.~\ref{fig:FHoverFInf} we plot the ratio between
the numerical fluxes at infinity and into the horizon $F^{\rm
  H}/F^{\infty}$ versus orbital velocity\footnote{Our $v \equiv
  (M\Omega)^{1/3}$ should not be confused with $v = (M/r)^{1/2}$ used
  in Ref.~\cite{Mino:1997bx}.  These definitions only agree when $q =
  0$.} for different values of the spin $q$. All curves in this figure
extend up to a point just outside their respective equatorial LRs; the
decreasing trend of $F^{\rm H}/F^{\infty}$ as a function of $q$ is
primarily due to how the factor $\Omega(\Omega - \Omega_{\rm H})$
behaves at the LR.  We indicate the position of the respective ISCOs
with vertical lines. For convenience, we list in
Table~\ref{tab:ISCOLR} the position of the ISCOs and LRs expressed in
terms of $v$ for the spin cases considered in this paper.
\begin{table*}
   \begin{ruledtabular}
    \begin{tabular}{c|ccccccccccc}%
      $q$ & $-0.99$ & $-0.9$ & $-0.7$ & $-0.5$  & 0 & 0.5 & 0.7 & 0.9 & 0.95 & 0.99\\ %
      \hline \\[-8pt]%
      $\; v_{\rm ISCO}\;$ &0.338 & 0.343 & 0.354 & 0.367 & 0.408 & 0.477 & 0.524 & 0.609 & 0.650 & 0.714 \\[2pt] %
      $\; v_{\rm LR}\;$ &0.523 & 0.527 & 0.536 & 0.546 & 0.577 & 0.625 & 0.655 & 0.706 & 0.729 & 0.763\\ %
    \end{tabular}
  \end{ruledtabular}
   \caption{\label{tab:ISCOLR} We show the orbital velocities corresponding to the positions of ISCO and LR for different values of the spin.}
\end{table*}

In Ref.~\cite{Yunes:2010zj} [see Fig.~2 therein] the authors
considered the total numerical flux $F^{\infty}_{\rm Teuk}+F^{\rm
  H}_{\rm Teuk}$ computed with the Teukolsky equation up to the ISCO
for different spins, and compared it to a flux model where
$F^{\infty}$ is the factorized flux of Ref.~\cite{Pan2010hz} and
$F^{\rm H}$ is the Taylor-expanded PN flux of
Refs.~\cite{Tagoshi:1997jy,Mino:1997bx}. They found that the inclusion
of the analytical ingoing flux is crucial for improving agreement with
the Teukolsky solution during the very long inspiral, implying that
$F^{\rm H}$ is a significant fraction of $F^{\infty}$. Our numerical
data extend the analysis of Ref.~\cite{Yunes:2010zj} to more extreme
spins (up to 0.99) and higher frequencies (up to the
LRs). Figure~\ref{fig:FHoverFInf} shows that $F^{\rm H}$ is typically
a few percent of $F^{\infty}$ at the ISCO for $q \leq 0.7$, increasing
to $8.7\%$ when $q = 0.99$.

Another important feature that Fig.~\ref{fig:FHoverFInf} shows is that
$F^{\rm H}$ changes sign for $q > 0$ ($F^{\infty}>0$ in all cases).
Orbits for which $F^{\rm H}/F^{\infty}<0$ are called ``superradiant.''
They can be interpreted as due to a Penrose-like
mechanism~\cite{Penrose:1969pc} in which the rotational energy of the
BH is extracted. The change of sign of $F^{\rm H}$ for $q> 0$ can be
understood by noticing that the sign of each mode $F^{\rm H}_{\ell m}$
is fixed by its specific structure in BH perturbation theory [see
Eq.~(\ref{alphalm})]
\begin{equation}
F^{\rm H}_{\ell m} = m^2\Omega\left(\Omega - \Omega_{\textrm{H}}\right) \tilde{F}^{\rm H}_{\ell m}\label{FHlmStructure}\,,
\end{equation}
where $\tilde{F}^{\rm H}_{\ell m} >0$. If $q>0$, $\Omega_{\rm H} >0$
as well, so when $0<\Omega< \Omega_{\textrm{H}}$, we have $F^{\rm
  H}_{\ell m} <0$.  This means that the particle gains energy through
the GW modes with that specific value of $m$.  Zeros in $F^{\rm H}$
for $q>0$ in Fig.~\ref{fig:FHoverFInf} coincide with the horizon
velocities: $v_{\rm H}\equiv (M \Omega_{\rm H})^{1/3}$. We notice that
for $q>0$, an inspiraling test particle will always go through the
zero of $F^{\rm H}$. In fact, the test-particle's velocity reaches its
maximum value, which is always larger than $v_{\rm H}$, during the
plunge. Afterwards, the test-particle's velocity decreases and gets
locked to that of the horizon~\cite{TaracchiniSpinTPL}.

As discussed in the Introduction and as can be seen in
Fig.\ \ref{fig:FHoverFInf}, we always have $|F^{\rm H}|/F^\infty < 1$,
meaning that we find no so-called ``floating orbits.''  Although
superradiance of the down-horizon modes does not allow for floating
orbits, these modes nonetheless have a strong impact on inspiral.
Comparing an inspiral that includes both $F^{\rm H}$ and $F^\infty$
with one that is driven only by $F^\infty$, one finds that these modes
make inspiral last longer, radiating additional cycles before the
final plunge~\cite{Yunes:2010zj}. A more quantitative assessment of
this delayed merger can be found for instance in the nonspinning limit
in Ref.~\cite{Bernuzzi:2012ku}.  In that work, the authors considered
EOB orbital evolutions which include the horizon flux model developed
in Ref.~\cite{Nagar:2011aa}. For $\mu/M=10^{-3}$, they found that
neglecting the horizon flux induces a dephasing of 1.6~rads for the
(2,2) mode waveform $h_{22}$ at merger over an evolution of about 41
orbital cycles.  They also studied what happens for larger mass
ratios, since their flux model worked even in the comparable-mass
limit. However, in this regime the effects are much smaller, with a
(2,2) mode dephasing of only $5\times 10^{-3}$~rads at merger
cumulated over 15 orbits. This result is consistent with the
estimations of Ref.~\cite{Alvi:2001mx}, which considered a
comparable-mass spinning case under a leading-order PN evolution.

In the case of spinning binaries with extreme mass-ratio,
Refs.~\cite{Hughes:1999bq,Hughes:2000ssa} found that in the nearly
extremal case $q=0.998$ the last few hundred days of inspiral at mass
ratio $10^{-6}$ are augmented by $\sim5\%$ at low inclinations,
depending on whether the ingoing flux is included or not. Using the
exact Teukolsky-equation fluxes of this paper in the EOB equations of
motion, Ref.~\cite{TaracchiniSpinTPL} (see Table~I therein) computed
how the number of orbital cycles within a fixed radial range before
the LR is affected by the addition of ingoing flux. Several different
values of the spin were considered. For prograde orbits, the ingoing
flux can increase the number of cycles by as much as $\sim7\%$ for
$q=0.99$, which corresponds to about 45~rads of GW dephasing in the
(2,2) mode over 100 GW cycles. On the other hand, for retrograde
orbits or nonspinning black holes, the horizon flux tends to make
inspiral faster, decreasing the number of cycles before plunge thanks
to the additional loss of energy absorbed by the horizon in these
cases.  The horizon flux changes the duration of inspiral by at most
$\sim1\%$ when $q=-0.99$, a somewhat less significant effect.

\begin{figure}[!ht]
  \begin{center}
    \includegraphics*[width=0.45\textwidth]{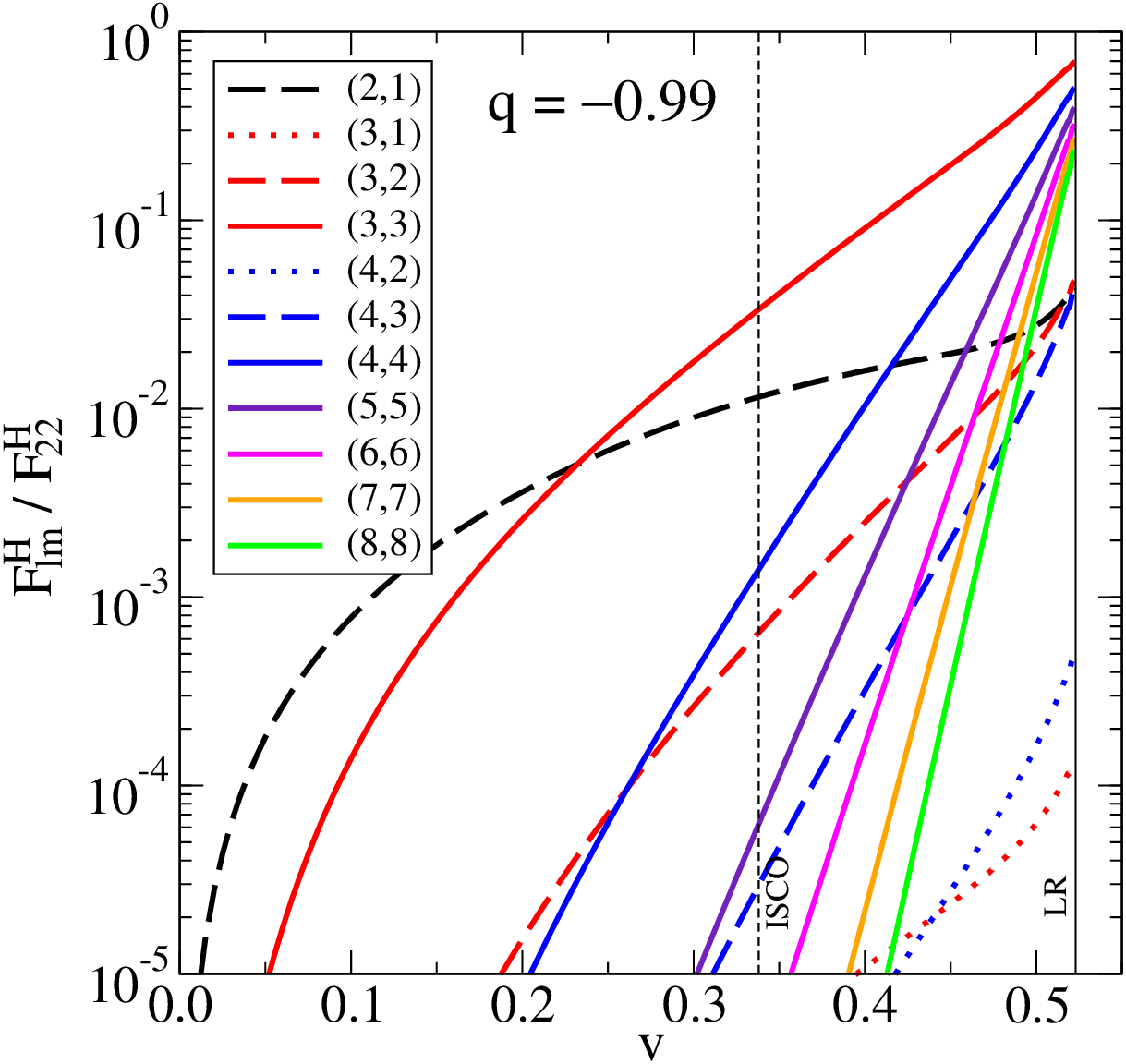}
    \caption{\label{fig:Hierarchym099} We compare the Teukolsky-equation ingoing multipolar fluxes, normalized by the dominant mode $F^{\rm H}_{22}$, for spin $q=-0.99$. Vertical lines mark the position of the ISCO and the LR. The graphs extend up to $r=r_{\rm LR}+0.01M$.}
  \end{center}
\end{figure}

\begin{figure}[!ht]
  \begin{center}
    \includegraphics*[width=0.475\textwidth]{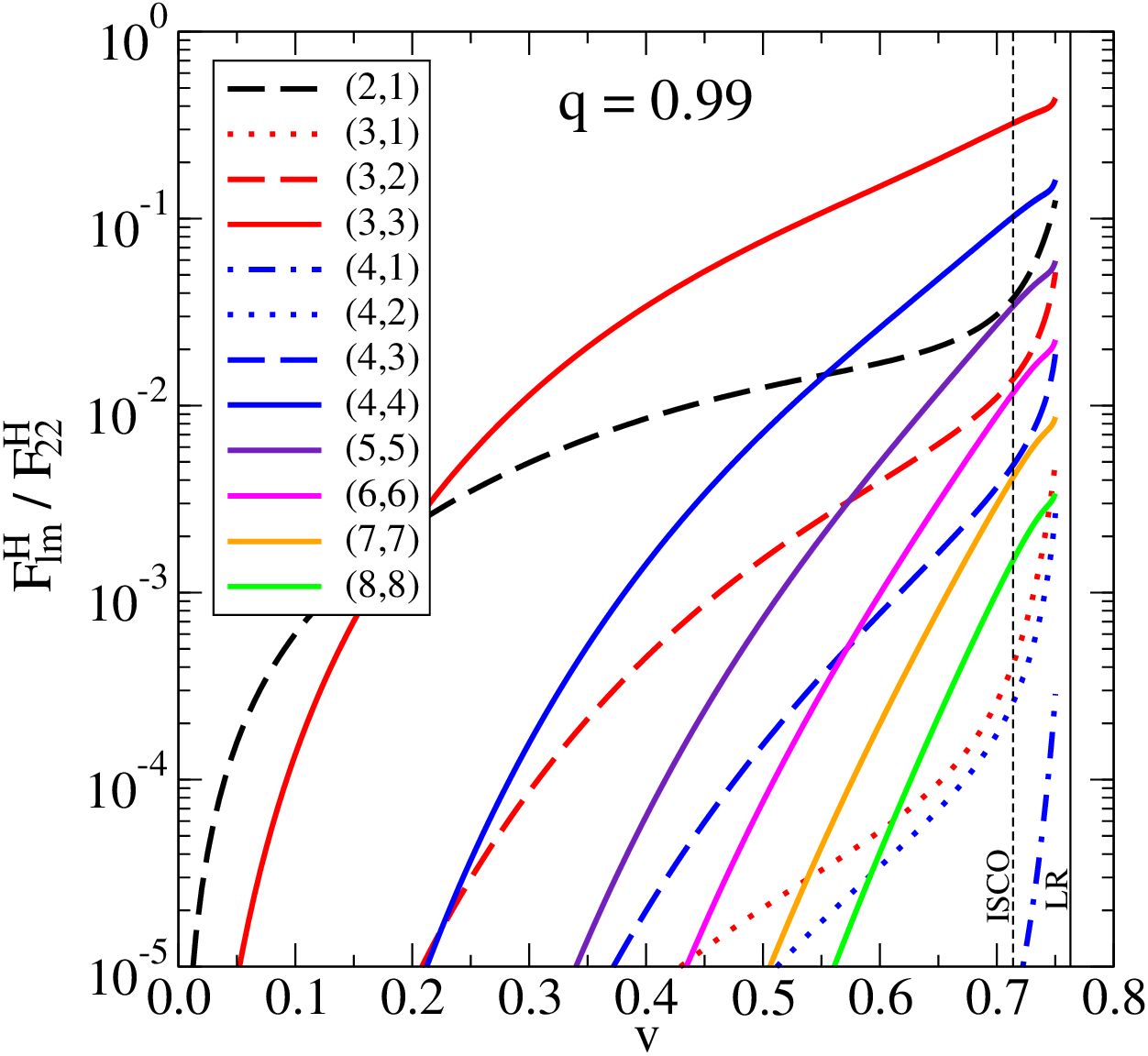}
    \caption{\label{fig:Hierarchy099} We compare the Teukolsky-equation ingoing multipolar fluxes, normalized by the dominant mode $F^{\rm H}_{22}$, for spin $q=0.99$. Vertical lines mark the position of the ISCO and the LR. The graphs extend up to $v\approx(M\Omega_{\rm H})^{1/3}$.}
  \end{center}
\end{figure}

Since we are going to model the multipolar modes $F^{\rm H}_{\ell m}$
rather than the total ingoing GW flux $F^{\rm H}$, it is useful to
understand their relative importance. In Figs.~\ref{fig:Hierarchym099}
and \ref{fig:Hierarchy099} we show the ratio between the first few
subdominant modes and the dominant (2,2) mode $F_{22}^{\rm H}$ as a
function of the orbital velocity for the two extremal spin cases
$q=\pm 0.99$. For $q=-0.99$ we note that at the ISCO the most
important subdominant modes are the (3,3) and the (2,1), and they are
both only a few percents of the dominant (2,2) mode. For $q=0.99$, at
the ISCO the subdominant modes which are at least $1\%$ of the (2,2)
mode are many more: (3,3), (4,4), (2,1), (5,5), (3,2) and (6,6). This
is a general result: as the spin of the Kerr BH grows to large
positive values, more and more multipolar modes become important
relative to the dominant (2,2) mode, even before the plunging phase,
which starts after the crossing of the ISCO. Close to the LR all modes
with $\ell=|m|$ become comparable to the (2,2) mode for both
spins. This is similar to what happens for the multipolar
decomposition of $F^{\infty}$ (see, e.g., Ref.~\cite{Finn:2000sy}).
Reference~\cite{Barausse:2011kb} already pointed out a similar
behavior while discussing the spherical modes at infinity $h_{lm}$,
which directly relate to the $-2$ spin-weighted spherical harmonic
decomposition of $F^{\infty}$ [see Eq.~(\ref{FInfModes}) below].

\begin{figure}[!ht]
  \begin{center}
    \includegraphics*[width=0.475\textwidth]{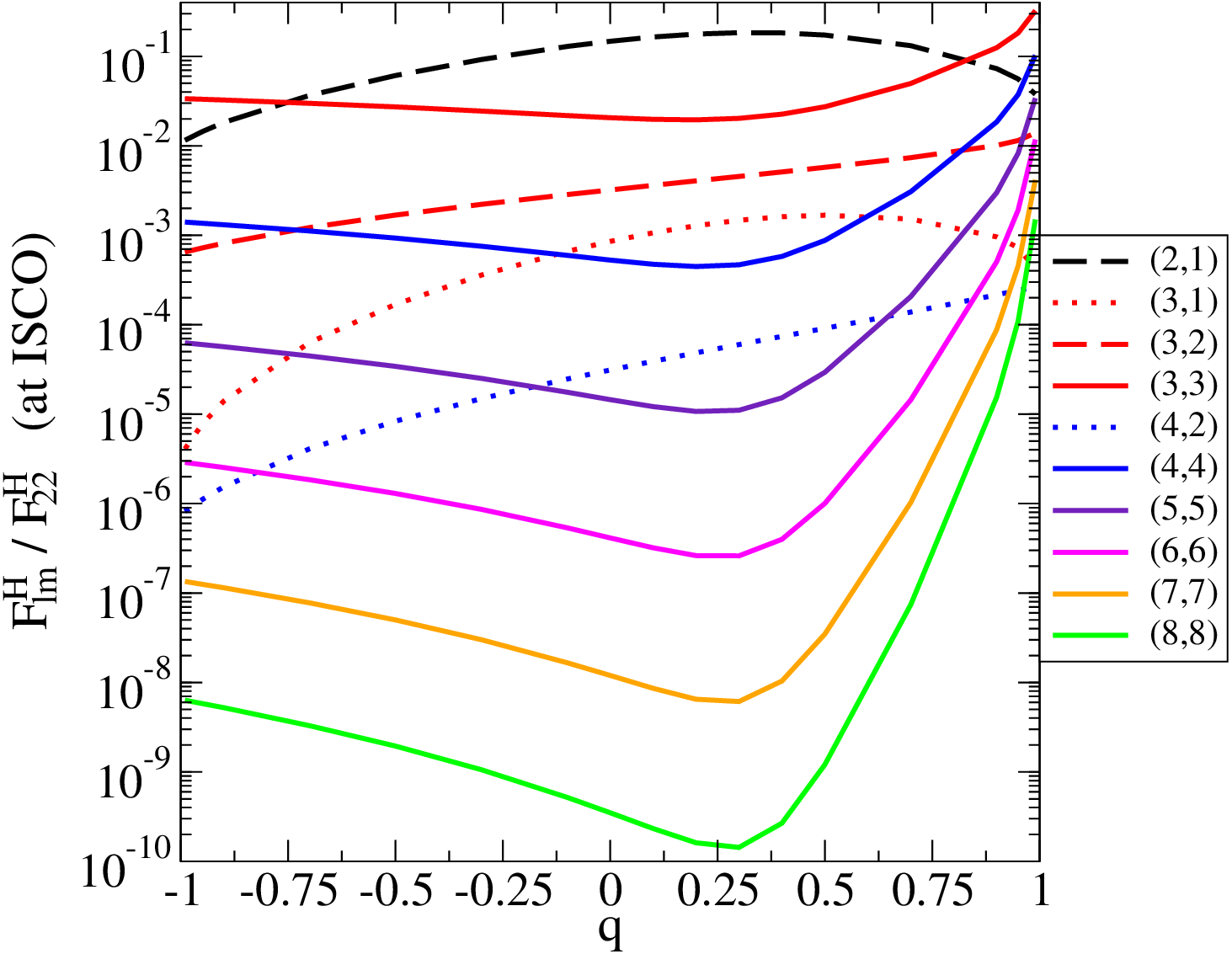}
    \caption{\label{fig:HierarchyISCO} We compare the Teukolsky-equation ingoing multipolar fluxes, normalized by the dominant mode $F^{\rm H}_{22}$, evaluated at the respective ISCOs.}
  \end{center}
\end{figure}

A compact representation of the ratio $F^{\rm H}_{\ell m}/F^{\rm
  H}_{22}$ across the entire range of physical spins is given in
Fig.~\ref{fig:HierarchyISCO}. Choosing to evaluate the ratio at the
same orbital frequency for different values of $q$ would not be
meaningful, since the position of the horizon changes with $q$, so we
choose instead as common physical point the ISCO for all the spins. We
see that at the ISCO the only modes which are consistently at least
$1\%$ of $F^{\rm H}_{22}$ are the (2,1) and (3,3) modes; only when $q
\gtrsim 0.95$ the (4,4), (5,5), (6,6) and (3,2) modes are above $1\%$
of the (2,2). Modes with $\ell=|m|$ appear to be evenly spaced on the
logarithmic scale used for all spins.  In other words, $F^{\rm
  H}_{\ell \ell}/F^{\rm H}_{22} \propto 10^{c(q) \ell}$, where $c(q)$
is a spin-dependent constant\footnote{This behavior is consistent with
  Eq.~(\ref{FellRatio}).}.  We therefore do not see crossings among
these modes as $q$ varies between $-1$ and 1. On the other hand, we do
see crossings between the largest subdominant modes, (2,1) and (3,3):
when $-0.75 \lesssim q \lesssim 0.8$ we have $F^{\rm H}_{21} \geq
F^{\rm H}_{33}$, otherwise (for almost extremal spins) $F^{\rm H}_{21}
\leq F^{\rm H}_{33}$. The nature of these crossings seems to depend
mostly on $|q|$, as it is also indirectly confirmed in
Figs.~\ref{fig:Hierarchym099} and \ref{fig:Hierarchy099}, where the
crossing of (2,1) and (3,3) (now considered in plots versus $v$ at
fixed $q$) occurs at a similar velocity $v\approx 0.2$ for both
$q=-0.99$ and $q=0.99$.  A simple explanation of what we just
discussed is the fact that, as $q$ grows, the ISCO moves deeper into
the strong field and the ISCO orbital velocity increases.  In this
circumstance, higher multipoles can become comparable in size to the
(2,2) mode in spite of their higher PN order.

From Figs.~\ref{fig:Hierarchym099}-\ref{fig:HierarchyISCO} we also
observe that, among modes with the same value of $\ell$, the dominant
ones are those with $\ell=|m|$, independently of the frequency. For the
case of scalar perturbations of a Schwarzschild BH,
Ref.~\cite{Breuer:1974uc} provided an analytical argument to account
for this peculiar hierarchy. Within the WKB approximation (valid for
$\ell \gg 1$) and for an orbit at $r \gg r_{\rm LR}$, it was shown
that $F^{\infty}_{\ell m}/F^{\infty}_{\ell
  \ell}\propto\exp{[-2C\,(\ell - |m|)]}$, where $C$ is a numerical
constant which depends on $r$.  As a consequence, nearly all of the
power at infinity at a frequency $m\Omega$ is emitted in the $\ell =
|m|$ modes. Similar arguments apply to the case of gravitational
perturbations~\cite{Breuer:1973kt} and, more generally, to
perturbations of a Kerr BH~\cite{Chrzanowski:1974nr}. Explicitly, one
finds that
\begin{equation} 
F^{\infty}_{\ell m} \propto \exp{\left[-2\int_{r_{\rm orb}^{*}}^{\bar{r}^{*}}{\sqrt{V(r'^{*})-m^{2}\Omega^{2}}\, 
\textrm{d}r'^{*}}\right]}\,.
\end{equation}
Here, $V$ is the radial potential seen by the perturbation, and $r^{*}$ is
the tortoise coordinate,
\begin{equation}
r^* = r + \frac{2Mr_+}{r_+ - r_-}\ln\left(\frac{r - r_+}{2M}\right)
- \frac{2Mr_-}{r_+ - r_-}\ln\left(\frac{r - r_-}{2M}\right)\;,
\end{equation}
where $r_\pm/M = 1 \pm \sqrt{1 - q^2}$.  The integral's upper limit
$\bar{r}^{*}$ is the larger of the two solutions to the equation
$V(\bar{r}^{*})=m^{2}\Omega^{2}$.  Recall that $\Omega$ depends on $r$
through Eq.~(\ref{KeplerKerr}). Note that $r_{\rm orb}^{*}$ is always
smaller than $\bar{r}^{*}$. For a nonspinning BH and $\ell \gg 1$ the
radial potential is the same regardless of the spin of the perturbing
field~\cite{Breuer:1974uc}, and reads $V(r) =
\ell(\ell+1)(1-2M/r)/r^{2}$. Therefore the lower the value of $m$, the
larger the value of $\bar{r}^{*}$, the larger the magnitude of the
argument inside the exponential, and hence the larger the
suppression. An analogous explanation applies to the absorption flux.

Finally, as we discussed in Sec.~\ref{sect:TeukSums}, the existence of
a cutoff value $\ell_{\rm c}$ for sums over the flux modes reduces in
practice the number of modes that contribute to the total flux. For
orbits very close to the LR, $\ell_{\rm c}$ is a decreasing function
of the spin. When $q\approx 1$ very few modes contribute, and the
total flux is basically given by the (2,2) mode. This is consistent
with Fig.~\ref{fig:Hierarchy099}, where in the strong-field region
only the (3,3), (4,4) and (2,1) modes are at least $10\%$ of the 
(2,2) mode. On the other hand, in Fig.~\ref{fig:Hierarchym099} we can see that the 
(3,3), (4,4), (5,5), (6,6), (7,7) and (8,8) modes are all larger than $10\%$ of the (2,2) mode 
at $r_{\rm min}$, and indeed the estimated $\ell_{\rm c}$ at that radial
separation is $\sim 200$.

\section{Factorization of the energy fluxes}
\label{sec:Fact}

The analytical representation of the ingoing flux in PN-expanded form
provided in Ref.~\cite{Mino:1997bx} turns out to be monotonic in the
orbital frequency for all possible values of the spin, so that the
sign-flip discussed above is not present. Moreover comparisons with
the numerical fluxes (see Fig.~\ref{fig:ErrTot}) show that these PN
formulae start performing poorly even before the ISCO, especially for
large positive values of $q$. This is to be expected, since the ISCO
moves to smaller radii (i.e. larger orbital frequencies) as $q$
increases, that is outside the range of validity of the PN
expansion. For instance, when $q=0.9$, the Taylor-expanded PN model for
$F^{\rm H}$ differs from the numerical data by more than 100\% around
an orbital velocity $v\approx 0.4$, while $v_{\textrm{ISCO}} \approx
0.61$. An improved analytical model for $F^{\rm H}$ is therefore
needed. In this section we will propose a factorization of the absorbed flux similar
to what was done for the flux at infinity \cite{Damour2007,DIN,
  Pan2010hz,Nagar:2011aa}. 

\subsection{Factorization of the energy flux at infinity}
\label{sec:InfFact}

For a particle spiraling in along an adiabatic sequence of circular
orbits, the GW flux at infinity can be expressed
as a sum over the waveform modes at infinity $h_{lm}$, as 
\begin{equation}
F^{\infty} = \frac{M^2 \Omega^2}{8\pi}\sum_{l = 2}^{\infty}\sum_{m=1}^{l} m^2 \bigg\lvert \frac{\mathcal{R}}{M} h_{lm} \bigg\rvert^2\,,\label{FInfModes}
\end{equation}
where $\mathcal{R}$ is the distance to the source.  The mode
decomposition here is done using the $-2$ spin-weighted spherical
harmonics, rather than the spheroidal harmonics considered in the
previous section; as discussed at the end of the introduction, the
indices are labeled $(l,m)$ rather than $(\ell, m)$ to flag this
change of basis. In Ref.~\cite{Damour2007} a novel approach to
improve the analytical modeling of the GW flux at infinity for a test
particle in Schwarzschild was introduced. This approach was then
generalized to spinning BHs in Ref.~\cite{Pan2010hz}. The idea is to
start from the PN knowledge of $h_{lm}$, and recast the formulae, mode
by mode, in a factorized form
\begin{equation}
h_{l m} \equiv h_{l m}^{(\textrm{N},\epsilon)} T_{l m}\hat{S}^{(\epsilon)}_{\textrm{eff}}  f_{l m} e^{i \delta_{l m}}\,,
\end{equation}
where $\epsilon$ is the parity of the $(l,m)$ mode, $h_{l
  m}^{(\textrm{N},\epsilon)}$ is the leading order term, $T_{l m}$ resums an
infinite number of leading logarithms entering the tail effects,
$\hat{S}^{(\epsilon)}_{\textrm{eff}}$ is an effective source term
which is divergent for circular motion at the LR, $f_{l m}$ and
$\delta_{l m}$ are polynomials in the variable $v$ [see, e.g., Ref.~\cite{Pan2010hz} for more details]. 
The term $f_{l m}$ is fixed by requiring that Eq.~(\ref{InfinityFactorization}), when expanded in
powers of $v$, agrees with the PN-expanded formulae. When computing
the $f_{l m}$'s, one assumes quasi-circular orbits, and this is
reflected by the choice of the source term,
\begin{equation}
\label{Seff}
  \hat{S}_\text{eff}^{(\epsilon)}=
  \begin{cases}
    \dfrac{E}{\mu}, &\textrm{if}\quad \epsilon = 0\,\,, \\
    \dfrac{L_{z}}{\mu M/  v}, &\textrm{if}\quad \epsilon = 1\,,
  \end{cases}
\end{equation}
where $E$ and $L_{z}$ are the energy and angular momentum of a circular equatorial orbit in Kerr~\cite{Bardeen:1972fi}
\begin{eqnarray}
\frac{E}{\mu} &=& \frac{1-2M/r + q (M/r)^{3/2}}{\sqrt{1 - 3M/r + 2 q (M/r)^{3/2}}}\, ,\label{Ecirc}\\
\frac{L_{z}}{\mu M} &=& \sqrt{\frac{r}{M}} \frac{1-2 q (M/r)^{3/2} + q^2 (M/r)^{2}}{\sqrt{1 - 3M/r + 2 q (M/r)^{3/2}}}\label{Lcirc}\, ,
\end{eqnarray}
and $\mu M /v$ in the denominator of Eq.~(\ref{Seff}) is the Newtonian
angular momentum for circular orbits. Note that this specific choice
of the effective source term is not the only one possible.  References
\cite{DIN,Pan2010hz} also explored the possibility of using
$\hat{S}_{\rm eff}^{(0)}=\hat{S}_{\rm eff}^{(1)}=E/\mu$, and labelled
the resulting factorized odd-parity modes with the ``H'' superscript
(meaning ``Hamiltonian''), as opposed to the factorization done with
the prescription in Eq.~(\ref{Seff}), whose odd-parity modes were
labelled with the ``L'' superscript (meaning ``angular momentum''). In
the rest of the paper we are going to consider only the effective
source of Eq.~(\ref{Seff}), and we will omit the ``L'' superscript.

Reference \cite{DIN} found that the 1PN coefficient of the $f_{lm}$
polynomials grows linearly with $l$, and therefore proposed a
better-behaved factorization, namely
\begin{equation}
h_{lm} \equiv h_{lm}^{(\textrm{N},\epsilon)} T_{lm}\hat{S}^{(\epsilon)}_{\textrm{eff}}  (\rho_{lm})^{l} e^{i \delta_{lm}}\,, \label{InfinityFactorization}
\end{equation}
where the $f_{lm}$ factor is replaced by $(\rho_{lm})^{l}$.  Both
factorized representations of $F^{\infty}$ show an improved agreement
with the numerical data with respect to PN approximants, as pointed
out in Refs.~\cite{Damour2007,DIN} for the nonspinning case and in
Ref.~\cite{Pan2010hz} for the spinning case. Moreover, the
$\rho_{lm}$--factorization turns out to perform better than the
$f_{lm}$--factorization when compared with the Teukolsky-equation
fluxes; this is discussed in more detail in Appendix
{\ref{sect:AppendixFitFInf}}.

\subsection{Factorization of the BH-absorption energy flux}
\label{sec:AbsFact}

Let us now consider the BH-absorption flux. For the special case of nonrotating BHs, 
Ref.~\cite{Poisson:1994yf} and Ref.~\cite{Taylor:2008xy} computed the lowest PN terms of $F^{\rm H}$,
in the test-particle and comparable-mass limit, respectively. The spinning case was considered in
Refs.~\cite{Tagoshi:1997jy,Mino:1997bx} in the test-particle limit and in 
Ref.~\cite{Alvi:2001mx} in the comparable-mass limit. In particular, Ref.~\cite{Mino:1997bx} 
computed the PN expanded BH-absorption flux into a Kerr BH up to 6.5PN order 
beyond the leading order luminosity at infinity for circular
orbits in the equatorial plane. The idea behind that calculation is to
solve the Teukolsky equation in two different limits, for separations
$r \rightarrow \infty$ and for separations approaching the horizon,
and then to match the two solutions in an intermediate region where
both approximations are valid. These Taylor-expanded PN expressions
are then decomposed into spheroidal multipolar modes $F^{\rm H}_{\ell
  m}$, so that
\begin{equation}
F^{\rm H}=2\sum_{\ell=2}^{\infty}\sum_{m=1}^{\ell}F^{\rm H}_{\ell m}\,,
\end{equation} 
where we used $F^{\rm H}_{\ell 0}=0$ and $F^{\rm
  H}_{\ell m}\equiv F^{\rm H}_{\ell |m|}$. Note that this decomposition stems from the separation 
of variables of the Teukolsky equation in oblate spheroidal coordinates~\cite{Teukolsky:1972my,
  Teukolsky:1974yv}.

Here, we count the PN orders with respect to the leading order luminosity at
infinity of Eq.~(\ref{FInfLO}), which can be rewritten
\begin{equation}
F^{\infty,\,\rm N}= \frac{32}{5}\left (\frac{\mu}{M}\right)^2\,v^{10}\,,
\end{equation}
for circular orbits. Thus, as discussed above, for a nonspinning
binary the leading order term in the BH-absorbed GW flux is 4PN
[$\mathcal{O}(v^8)$ beyond the leading order luminosity at infinity],
whereas for a Kerr BH it is 2.5PN [$\mathcal{O}(v^5)$ beyond the
  leading order luminosity at infinity].

Reference~\cite{Nagar:2011aa} considered the case of a nonspinning BH
binary and applied a factorization to the multipolar ingoing GW
flux, recasting it in
the following form
\begin{equation}
F^{\rm H}_{\ell m}\equiv F^{\rm H, \, N}_{\ell m} (\hat{S}_{\textrm{eff}}^{(\epsilon)})^2 \left(\rho^{\rm H}_{\ell m}\right)^{2\ell}\, ,\label{BHAFactorization}
\end{equation}
where $F^{\rm H, \, N}_{\ell m}$ is the nonspinning leading term, and
$\rho_{\ell m}^{\rm H}$ is a polynomial in $v$ determined by requiring
that Eq.~(\ref{BHAFactorization}) agrees with the PN-expanded formulae
from Refs.~\cite{Poisson:1994yf,Taylor:2008xy} when expanded in powers
of $v$. Here the ``H'' superscript refers to ``horizon.'' Note that
Ref.~\cite{Nagar:2011aa} defined the multipolar modes differently:
their $(\ell,m)$ mode is the sum of our $(\ell,m)$ and $(\ell,-m)$
modes, so there is an overall factor $1/2$.
Reference~\cite{Nagar:2011aa} computed $\rho_{22}^{\rm H}$ up to 1PN
order beyond $F_{22}^{\rm H,N}$ (i.e., 5PN order in our convention) in
the Schwarzschild case and also in the comparable-mass case. However,
in the Schwarzschild case, the total ingoing GW flux is actually known
through 6PN order~\cite{Mino:1997bx}
\begin{equation}
F^{\rm H}(q=0) = F^{\infty,\,\rm N} v^{8}\left[1+4v^2+\frac{172}{7}v^4+\mathcal{O}(v^5)\right]\,,\label{NSTaylorFlux}
\end{equation}
and specifically the individual mode $F_{22}^{\rm H}$ is known to the same PN order as $F^{\rm H}$, so that the factorization in Ref.~\cite{Nagar:2011aa} can be extended from 5PN to 6PN order (beyond the leading order luminosity at 
infinity).

Let us now consider the spinning case. As pointed out before, the
Taylor-expanded PN form of the ingoing GW flux does not preserve the
zero $(\Omega - \Omega_{\textrm{H}})$, which is instead present in the
exact expression of the $F^{\rm H}_{\ell m}$'s from BH perturbation
theory. This means that, if we were to use a factorization like the
one in Eq.~(\ref{BHAFactorization}) also for the Kerr case, our
factorized flux would inherit this unwanted feature, since the
factorization only tries to match the Taylor-expanded PN
flux. Therefore, we propose the factorized form
\begin{equation}
F_{\ell m}^{\rm H} \equiv  \left(1 - \frac{\Omega}{\Omega_{\textrm{H}}}\right)F^{\rm H,\, N}_{\ell m} (\hat{S}_{\textrm{eff}}^{(\epsilon)})^2 (\tilde{f}^{\rm H}_{\ell m})^{2}\, ,\label{BHAfFactorizationTilde}
\end{equation}
which has the advantage of enforcing the presence of the zero at a frequency equal to $\Omega_{\textrm{H}}$. 
The leading term is defined as
\begin{equation}
F^{\rm H,\, N}_{\ell m}  \equiv \frac{32}{5}\left (\frac{\mu}{M}\right )^2\,v^{7 + 4\ell +2\epsilon}n^{(\epsilon)}_{\ell m}c_{\ell m}(q)\,,
\end{equation}
where 
\begin{eqnarray}
n^{(0)}_{\ell m} &\equiv& -\frac{5}{32}\frac{(\ell+1)(\ell+2)}{\ell(\ell-1)}\frac{2\ell+1}{[(2\ell+1)!!]^2}\times\notag\\
&\times&\frac{(\ell-m)!}{[(\ell-m)!!]^2} \frac{(\ell+m)!}{[(\ell+m)!!]^2} \,,
\end{eqnarray}
\begin{eqnarray}
n^{(1)}_{\ell m} &\equiv& -\frac{5}{8\ell^2}\frac{(\ell+1)(\ell+2)}{\ell(\ell-1)}\frac{2\ell+1}{[(2\ell+1)!!]^2}\times\notag\\
&\times&\frac{[(\ell-m)!!]^2}{(\ell-m)!}\frac{[(\ell+m)!!]^2}{(\ell+m)!} \,,
\end{eqnarray}
and
\begin{eqnarray}
c_{\ell m}(q) &\equiv& \frac{1}{q}\prod_{k=0}^{\ell}{\left[k^2 + \left(m^2 - k^2\right) q^2\right]}=qm^2\left(1 - q^2\right)^{\ell}\times\notag\\
&\times& \left(1 - \frac{i m q}{\sqrt{1 - q^2}}\right)_{\ell} \left(1 + \frac{i m q}{\sqrt{1 - q^2}}\right)_{\ell} \,,
\end{eqnarray}
where $(z)_n\equiv z(z-1)\cdots(z-n+1)$ is the Pochhammer symbol. The
factors $n^{(\epsilon)}_{\ell m}$ and $c_{\ell m}(q)$ allow the
$\tilde{f}^{\rm H}_{\ell m}$'s to start with either 1 or 0. The
definition of the factor $c_{\ell m}(q)$ is inspired by the derivation
of the $\ell=2$ modes in the slow-motion approximation in
Ref.~\cite{Poisson:2004cw} [see Eq.~(9.31) therein]. The definition of $n^{(\epsilon)}_{\ell m}$ is derived from
Eqs.~(5.17) and (5.18) in Ref.~\cite{Poisson:1994yf} (which considered
the Schwarzschild case), but a few additional factors included.  These
new factors are a prefactor of $1/(m\ell !)^2$ generated by our
definition of $c_{\ell m}(q)$; a numerical factor of $-1/4$ due to the
presence of $(1-\Omega/\Omega_{\rm H})$ in
Eq.~(\ref{BHAfFactorizationTilde}); and a factor of $1/2$ due to the
definitions used in Ref.~\cite{Poisson:1994yf}.  We also consider the
factorization
\begin{equation}
F_{\ell m}^{\rm H} \equiv  \left(1 - \frac{\Omega}{\Omega_{\textrm{H}}}\right)F^{\rm H,\, N}_{\ell m} (\hat{S}_{\textrm{eff}}^{(\epsilon)})^2 \left(\tilde{\rho}^{\rm H}_{\ell m}\right)^{2\ell}\, ,\label{BHAFactorizationTilde}
\end{equation}
where the factor $\tilde{f}_{\ell m}^{\rm H}$ in Eq.~(\ref{BHAfFactorizationTilde}) is replaced by
$\left(\tilde{\rho}^{\rm H}_{\ell m}\right)^{\ell}$, just as was
done by Ref.~\cite{DIN} for $F^{\infty}$. [Note that our
  $\tilde{\rho}_{\ell m}$'s are different from the $\rho_{\ell m}$'s in Ref.~\cite{Nagar:2011aa}.]

Appendix I of Ref.~\cite{Mino:1997bx} lists the Taylor-expanded modes
$F^{\rm H}_{\ell m}$ that are needed to compute the BH-absorption
Taylor-expanded flux through 6.5PN order. Since the $F^{\rm H}_{\ell
  m}$'s in Ref.~\cite{Mino:1997bx} are expressed in terms of the
velocity parameter $(M/r)^{1/2}$ we use Eq.~(\ref{KeplerKerr}) to
replace $r$ with $v$.  A straightforward but tedious calculation gives
us the following expressions for the $\tilde{\rho}^{\rm H}_{\ell m}$
functions:
\begin{widetext}
\begin{subequations}
\label{rho2}
\begin{align}
 \tilde{\rho}^{\rm H}_{22}&=1 + v^2 - \bigg\{2 B_2 + \frac{q}{1+3q^2} \left[4 +  \kappa \left(5  + 3 q^2\right)\right]\bigg\}v^3 +  \left(\frac{335}{84} -\frac{2}{21} q^2\right) v^4 \notag \\
&-\bigg\{2B_2+\frac{q}{1+3q^2} \left[\frac{47}{18} -\frac{25}{6} q^2 +\kappa \left(5+ 3q^2\right) \right]\bigg\} v^5 + \bigg\{\frac{293\,243}{14\,700} - \frac{2}{3}\pi^2 -\frac{6\,889}{1\,134}q^2+\frac{3}{2}q^4+2B_2^2\notag\\
&+4C_2 \bigg(1+\frac{2}{\kappa}\bigg) -\frac{428}{105}\left(A_2+\gamma_E+\log{2}+\log{\kappa}+2\log{v}\right)-\frac{1}{1+3q^2}\left[\frac{124}{9}-8 q B_2-2q \kappa B_2\left(5+3q^2\right)\right]\notag\\
&+\frac{1}{\left(1+3q^2\right)^2}\left[\frac{56}{3}+2\kappa\left(5-6q^2+3q^4-18q^6\right)\right]\bigg\} v^6 - \frac{1}{42}\bigg\{B_2\left(335-8q^2\right)+\frac{q}{1+3q^2}\bigg[\frac{1\,670}{3} - \frac{3\,131}{9} q^2 + \frac{73}{3} q^4\notag\\
&+\frac{\kappa}{2}\left(5+3q^2\right)\left(335-8q^2\right)\bigg]\bigg\}v^7+ \bigg\{\frac{6\,260\,459}{151\,200} - \frac{2}{3} \pi^2 -\frac{25\,234}{5\,292}q^2+\frac{8\,439}{5\,292}q^4-\frac{148}{7}\gamma_E-\frac{428}{105}A_2+2B_2^2 \notag\\
&+4C_2\left(1+\frac{2}{\kappa}\right)-\frac{25}{9}qB_2 + \frac{1}{1+3q^2}\left[-\frac{322}{27}+8qB_2+2\kappa q B_2\left(5+3q^2\right)\right]+ \frac{1}{\left(1+3q^2\right)^2}\bigg[\frac{56}{3}+\kappa \bigg(10 - \frac{341}{18} q^2 \notag\\ 
& - 19 q^4- \frac{97}{2} q^6\bigg)\bigg]-\frac{4\,012}{105}\log{2}-\frac{428}{105}\log{\kappa}-\frac{2\,648}{105}\log{v}\bigg\} v^8+ \mathcal{O}(v^9)\,, \label{rho22}\\
\tilde{\rho}^{\rm H}_{21}&=1-\frac{q}{3} v+\left(\frac{7}{12}-\frac{q^2}{18}\right)v^2- \bigg\{B_1 +\frac{1}{18}q \left(\frac{1}{3}q^2-\frac{31}{2}\right)+ \frac{q}{4 - 3 q^2}\left[1+\kappa \left(5-3q^2\right)\right]\bigg\}v^3+\bigg\{\frac{521}{672}+\frac{1}{3}q B_1\notag\\
&-q^2 \left(\frac{1\,847}{1\,512} + \frac{5}{648}q^2\right)+\frac{1}{4-3q^2}\bigg[\frac{4}{9}+q^2\kappa\bigg(\frac{5}{3}-q^2\bigg)\bigg]\bigg\}v^4+\bigg[-\frac{B_1}{36}\left(21-2q^2\right)-\frac{1}{4-3q^2}\bigg(-\frac{347}{72} q \notag\\
&+ \frac{3\,053}{864} q^3 + \frac{703}{1\,944} q^5 - \frac{7}{648} q^7+ \frac{1}{36} \kappa q \left(21-2q^2\right)\left(5-3q^2\right)\bigg)\bigg]v^5 + \bigg\{\frac{267\, 092\, 969}{38\, 102\, 400}-\frac{32\, 125}{12\, 096}q^2+\frac{81\, 167}{54\, 432}q^4\notag\\
&-\frac{7}{3\, 888}q^6-\frac{107}{105}\left(A_1+\gamma_E+\log{2}+\log{\kappa}+2\log{v}\right)+\frac{1}{2} B_1^2+C_1\bigg(1+\frac{2}{\kappa}\bigg)-\frac{\pi^2}{6}-\frac{1}{4-3q^2} \bigg[\frac{298}{243}\notag\\
&+ q B_1\bigg(\frac{22}{9} -\frac{287}{108} q^2 + \frac{1}{18}q^4-\kappa \left(5-3q^2\right)\bigg)\bigg]+\frac{1}{\left(4-3q^2\right)^2}\bigg[-\frac{4}{3}+\kappa\bigg(40-\frac{1\, 208}{9}q^2 + \frac{14\, 539}{108}q^4 \notag\\
&- \frac{177}{4}q^6 +\frac{1}{6}q^8\bigg)\bigg]\bigg\}v^6+ \mathcal{O}(v^7)\label{rho21} \,,
\end{align}
\end{subequations}
\begin{subequations}
\label{rho3}
\begin{align}
\tilde{\rho}^{\rm H}_{33}&=1+\frac{7}{6}v^2 -\bigg\{2B_3+\frac{2q}{\left(1+8q^2\right)\left(4+5q^2\right)}\bigg[\frac{131}{9}+\frac{314}{9}q^2-\frac{40}{9}q^4 + 3\kappa\left(5+13q^2\right)\bigg] \bigg\} v^3\notag\\
&+\bigg(\frac{353}{120} - \frac{5}{18} q^2\bigg)  v^4+\mathcal{O}(v^5)\,,\label{rho33}\\
\tilde{\rho}^{\rm H}_{32}&=1-\frac{1}{4}q v+\bigg(\frac{5}{6}-\frac{1}{16}q^2\bigg) v^2+\mathcal{O}(v^3)\label{rho32} \,,\\
\tilde{\rho}^{\rm H}_{31}&=1+ \frac{29}{18}v^2 -\frac{2}{3}\bigg\{B_1 + \frac{q}{4-3q^2}\bigg[\kappa \left(5-3q^2\right)+\frac{1}{9-8q^2}\bigg(65-\frac{866}{9}q^2+\frac{104}{3}q^4\bigg)\bigg]\bigg\}v^3 \notag\\
&+\bigg(\frac{1\, 903}{648}+\frac{1}{6}q^2\bigg)v^4+\mathcal{O}(v^5)\label{rho31} \,,
\end{align}
\end{subequations}
\begin{subequations}
\label{rho4}
\begin{align}
\tilde{\rho}^{\rm H}_{44}&=1+\mathcal{O}(v)\,,\label{rho44}\\
\tilde{\rho}^{\rm H}_{43}&=\mathcal{O}(v)\label{rho43} \,,\\
\tilde{\rho}^{\rm H}_{42}&= 1+\mathcal{O}(v)\label{rho42} \,,\\
\tilde{\rho}^{\rm H}_{41}&=\mathcal{O}(v)\label{rho41} \,.
\end{align}
\end{subequations}
\end{widetext}
In these equations, $\gamma_E\approx 0.57721\dots$ is the
Euler-Mascheroni constant, $\kappa \equiv \sqrt{1-q^2}$, and
\begin{eqnarray}
A_n &\equiv& \frac{1}{2}\left[\psi^{(0)}\left(3+\frac{i nq}{\kappa}\right) + \psi^{(0)}\left(3-\frac{i nq}{\kappa}\right)\right]\,,\\
B_n &\equiv& \frac{1}{2i}\left[\psi^{(0)}\left(3+\frac{i nq}{\kappa}\right) - \psi^{(0)}\left(3-\frac{i nq}{\kappa}\right)\right]\,,\\
C_n &\equiv& \frac{1}{2}\left[\psi^{(1)}\left(3+\frac{i nq}{\kappa}\right) + \psi^{(1)}\left(3-\frac{i nq}{\kappa}\right)\right]\,;
\end{eqnarray}
$\psi^{(n)}$ is the polygamma function. 
\begin{figure}[!ht]
  \begin{center}
    \includegraphics*[width=0.45\textwidth]{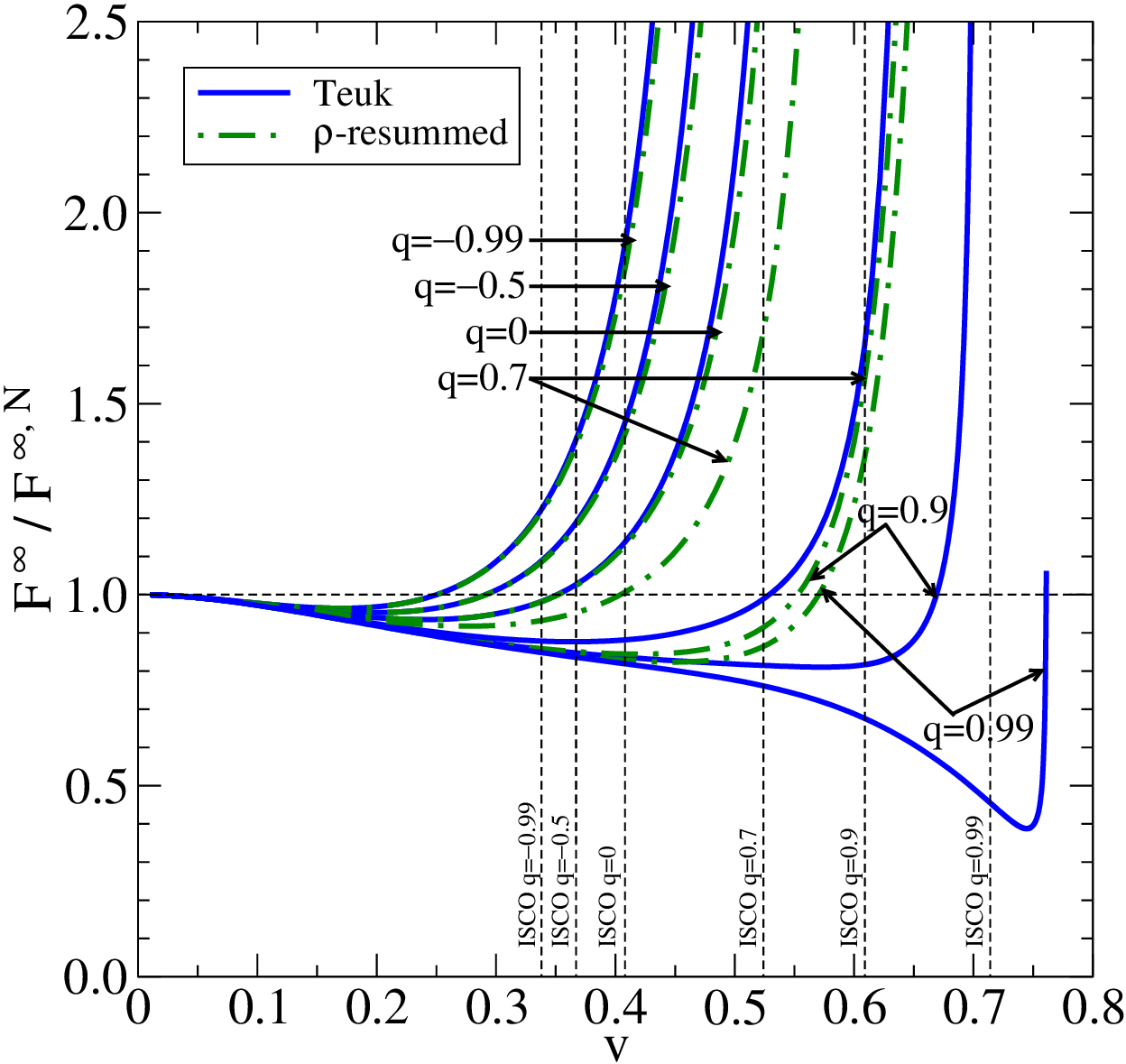}
    \caption{\label{fig:FInfLR} We compare the Teukolsky-equation flux at
      infinity with the factorized flux of Ref.~\cite{Pan2010hz}. The
      computation is done up to the $r_{\rm LR}+0.01M$.}
    \end{center}
\end{figure}

The explicit expressions of the $\tilde{f}^{\rm H}_{\ell m}$ functions can be found in
Appendix~\ref{sect:Appendixflm}. Given the limited number of available
modes in Taylor-expanded PN form, we are not able to convincingly argue 
that the $\tilde{\rho}^{\rm H}_{\ell m}$--factorization is preferable to 
the $\tilde{f}^{\rm H}_{\ell m}$--factorization on
the basis of the growth with $\ell$ of the 1PN coefficient in the $\tilde{f}^{\rm H}_{\ell m}$'s,  
as done in Refs.~\cite{DIN, Pan2010hz} for $F^{\infty}$. We prefer the $\tilde{\rho}^{\rm H}_{\ell
  m}$--factorization over the $\tilde{f}^{\rm H}_{\ell
  m}$--factorization because we find that it compares better to the numerical data.

\begin{figure*}
  \centerline{
    \includegraphics*[scale=0.4]{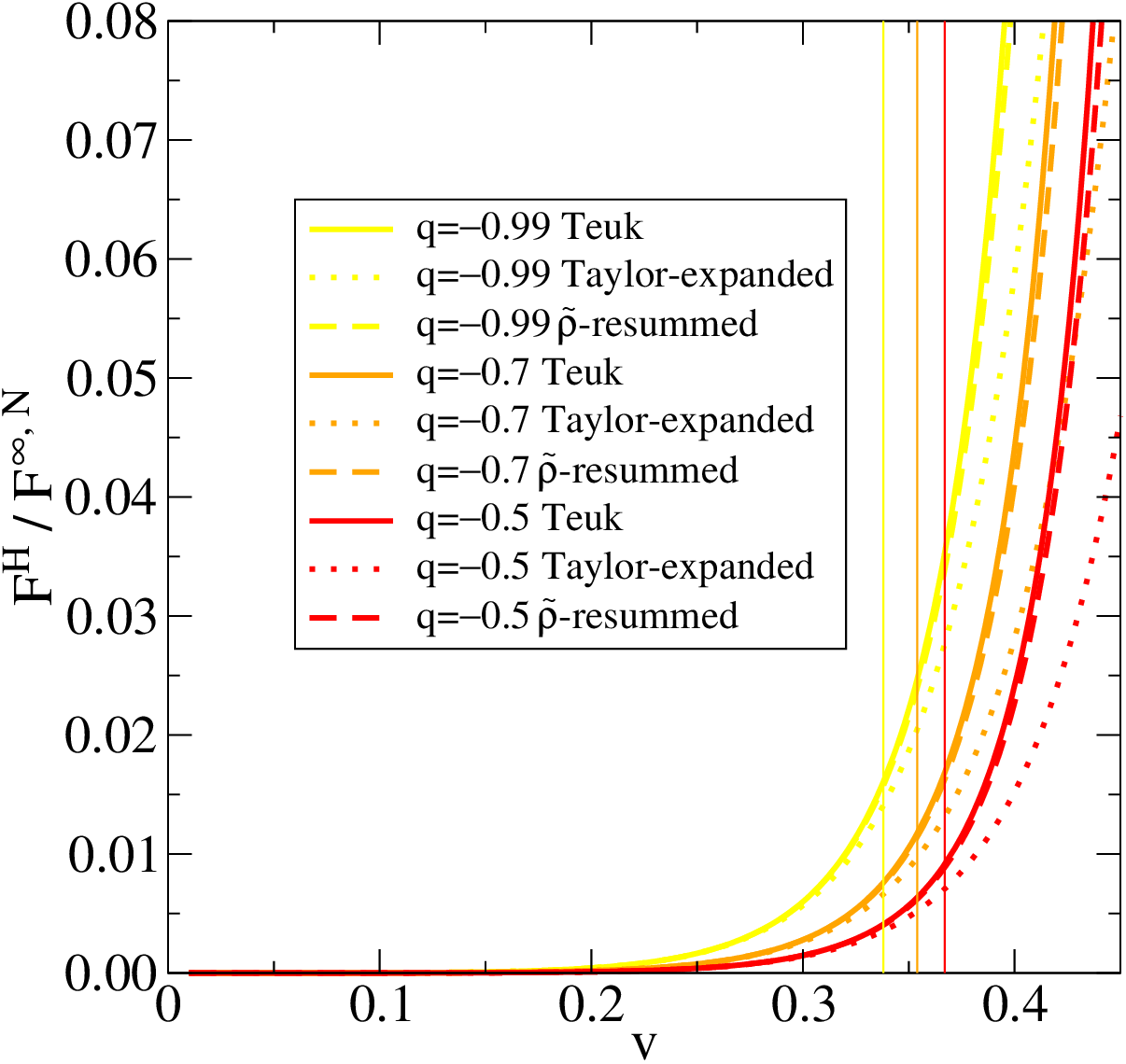}
    \hspace*{2em}
    \includegraphics*[scale=0.4]{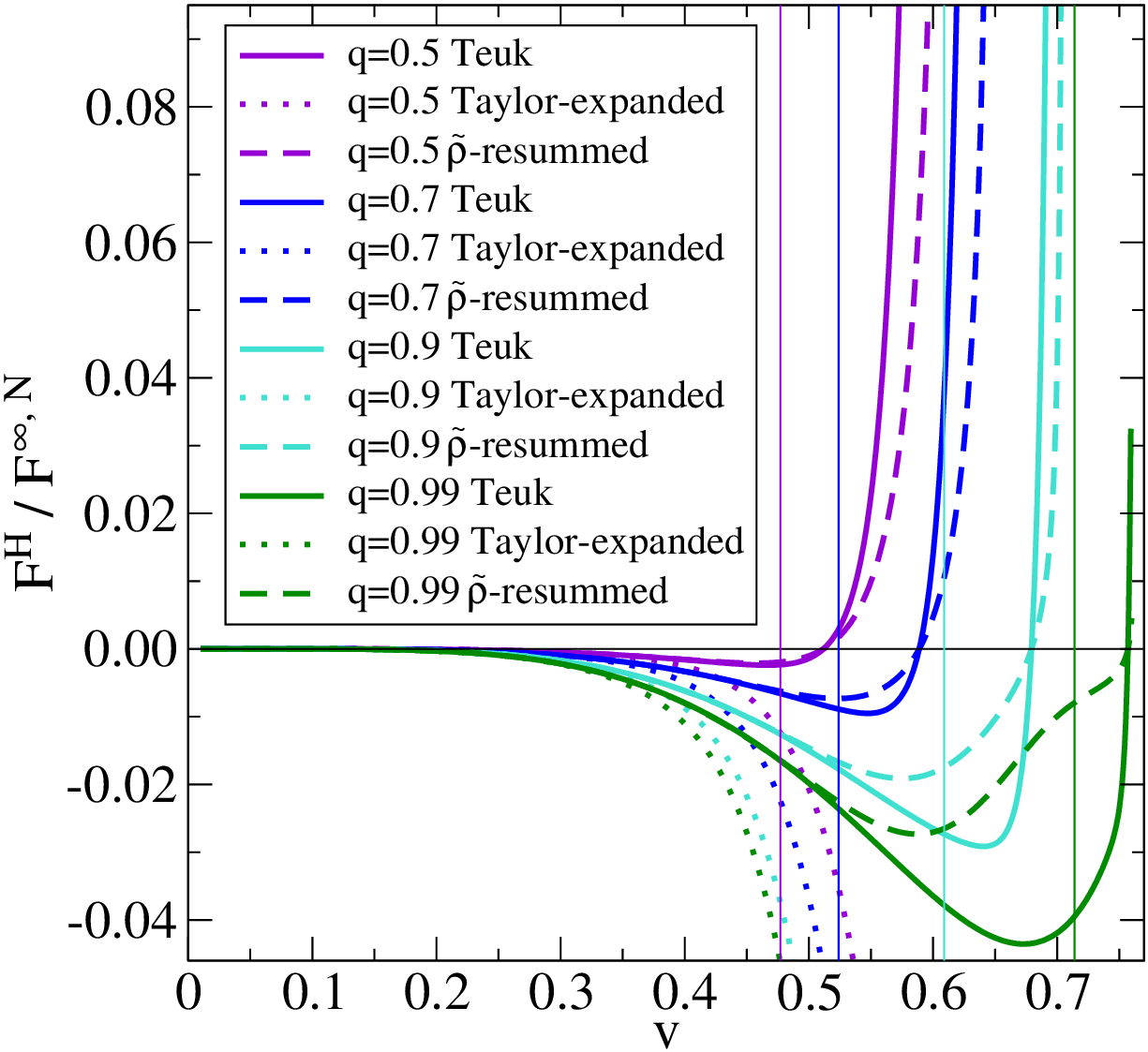}} 
 \caption{\label{fig:FHTot} We compare the Teukolsky-equation BH-absorption flux (solid lines) to the Taylor-expanded PN model of Ref.~\cite{Mino:1997bx} (dotted lines) and the factorized flux proposed in this work (dashed lines), as functions of $v$. All curves extend up to $r=r_{\rm LR}+0.01M$. Vertical lines mark the positions of the respective ISCOs. The fluxes are normalized to the leading order flux at infinity $F^{\infty,\rm N}$. In the left panel we show cases with $q<0$, while in the right panel we show cases with $q>0$.}
\end{figure*}

\section{Comparison with numerical results}
\label{sec:Comparisons}

In this section we compare the Teukolsky-equation fluxes (both at infinity and
ingoing) to the analytical models discussed in Sec.~\ref{sec:Fact}.

\subsection{Comparison with the numerical flux at infinity}

In Fig.~\ref{fig:FInfLR} we show the Teukolsky-equation flux at
infinity for several different spin values up to the LR and compare it
to the factorized flux reviewed in Sec.~\ref{sec:InfFact} and
developed in Ref.~\cite{Pan2010hz}. We note that the factorized flux
is fairly close to the numerical data until the LR for retrograde and
nonspinning cases.  For large spin prograde cases, the modeling error
instead becomes large already at the ISCO\footnote{Besides the
  $\rho_{\ell m}$--factorization discussed in Sec.~\ref{sec:InfFact},
  Ref.~\cite{Pan2010hz} also proposed an improved resummation of the
  $\rho_{\ell m}$ polynomials, which consists in factoring out their
  0.5PN, 1PN and 1.5PN order terms, with a significant improvement in
  the modeling error.}. Following the approach of
Ref.~\cite{Yunes:2010zj}, in Appendix~\ref{sect:AppendixFitFInf} we
have improved the factorized flux at infinity by fitting the
$\rho_{\ell m}$'s to the Teukolsky-equation data. These fits can be
useful for very accurate numerical evolution of PN or EOB equations of
motions for EMRIs, and also for the merger modeling of small
mass-ratio binary systems~\cite{TaracchiniSpinTPL}.

\subsection{Comparison with the numerical flux through the black-hole horizon}

\begin{figure}[!ht]
  \begin{center}
    \includegraphics*[width=0.45\textwidth]{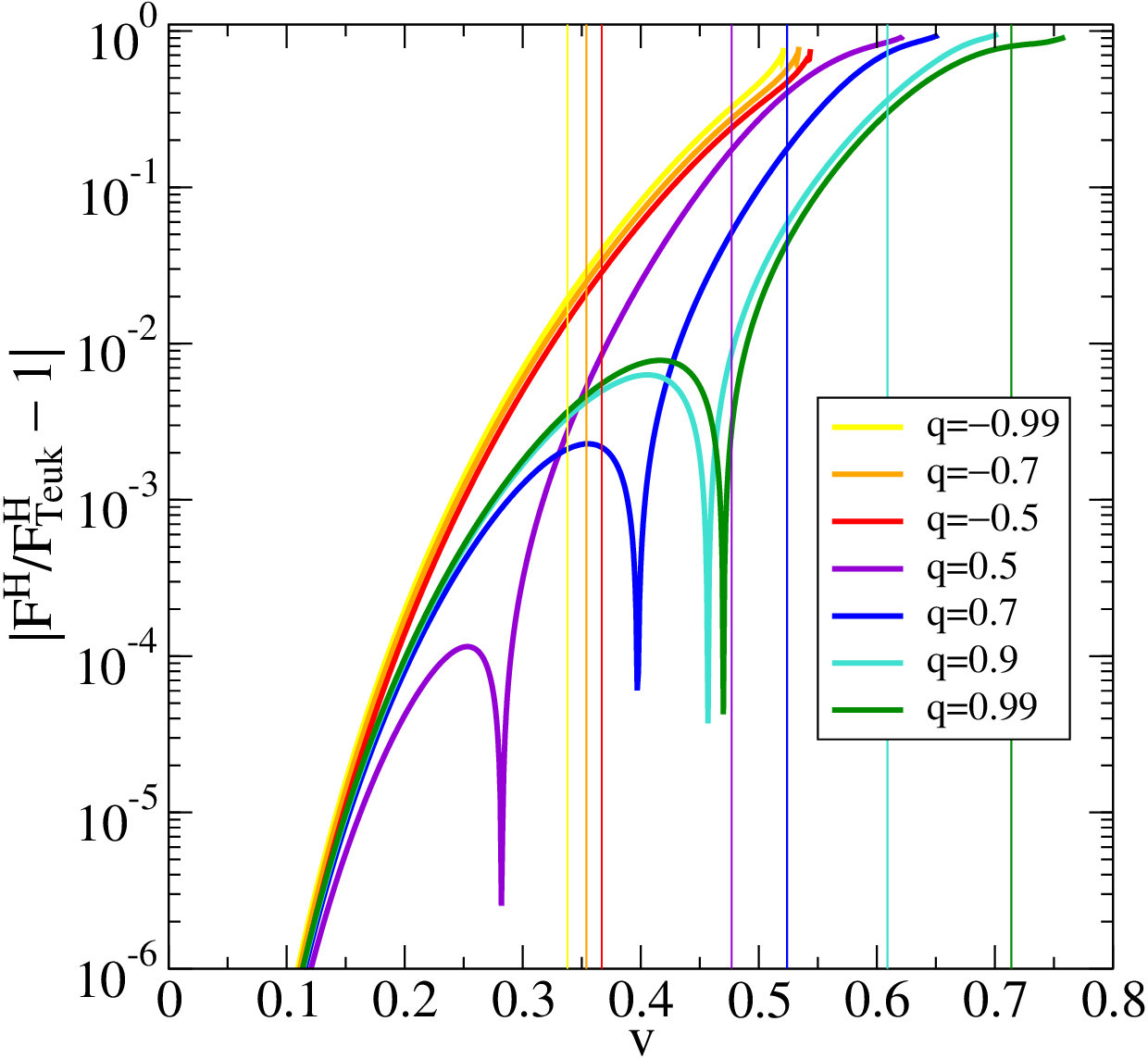}
    \caption{\label{fig:ErrTot} We show the fractional difference between 
      the total factorized and Teukolsky-equation fluxes. All curves 
      extend up to the respective LRs. Vertical lines mark the
      positions of the respective ISCOs.}
  \end{center}
\end{figure}

In Fig.~\ref{fig:FHTot} we compare the BH-absorption Taylor-expanded
PN flux from Ref.~\cite{Mino:1997bx} and our factorized flux to the
numerical flux produced with the frequency-domain Teukolsky equation,
normalized to the leading order luminosity at infinity. In
Fig.~\ref{fig:ErrTot} we plot the fractional difference between
numerical and factorized fluxes. The factorized model is quite
effective in reproducing the numerical data, not only because we have
factorized the zero $(1 - \Omega/\Omega_{H})$ in
Eq.~(\ref{BHAfFactorizationTilde}), but also because we have
factorized the pole at the LR through the source term
$\hat{S}_{\textrm{eff}}^{(\epsilon)}$ in
Eq.~(\ref{BHAfFactorizationTilde}). As we see in Fig.~\ref{fig:FHTot},
the factorized flux is quite close to the numerical flux up to $q\leq
0.5$, but starts performing not very well soon after the ISCO when
$q\geq0.7$, systematically underestimating $|F^{\rm H}|$ in the range
$v_{\rm ISCO} < v < v_{\rm H}$ for large positive spins. As we see in
Fig.~\ref{fig:ErrTot}, for spins $-1 \le q \le 0.5$ the agreement of
the factorized model to the numerical data is better than $1\%$ up to
the ISCO, with a remarkable improvement over the Taylor-expanded PN
model. For instance, for $q=0.5$, the ISCO is located at
$v_{\textrm{ISCO}}\approx 0.48$.  Up to the ISCO the agreement is
below $1\%$, while in the last part of the frequency range (up to the
LR) we see that the performance becomes worse. For larger spins the
factorized model starts to visibly depart from the numerical data even
before the ISCO, but the error is still within $50\%$ at the ISCO for
$q=0.9$. By contrast, the Taylor-expanded PN model is completely
off. For positive spins we see that the relative error of the
factorized model goes to zero at $v=(M \Omega_{\rm H})^{1/3}\equiv
v_{\rm H}$, which is where our model by construction agrees with the
Teukolsky-equation data thanks to the factor $(1-\Omega/\Omega_{\rm
  H})$. On the other hand, the Taylor-expanded PN model has the wrong
sign at high frequencies when $q>0$.

The large modeling error of the factorized flux for $q \geq 0.7$ after
the ISCO should not be a reason for significant concern.  Physical
inspirals will not include circular motion beyond the ISCO; the main
purpose of modeling fluxes from these orbits is to properly include
the influence of this pole near the light ring.  The physical motion
will in fact transition to a rapid plunge near the ISCO, generating
negligible flux.  In Ref.~\cite{TaracchiniSpinTPL}, we evolved EOB
equations of motions incorporating the absorption flux into the
radiation reaction force. We found that using the exact
Teukolsky-equation flux or the factorized model flux of this paper
makes very little difference in terms of the duration of the
inspiral. For the large spin cases (i.e., those with the largest
modeling error even before the ISCO) the length of the inspiral varies
by at most $\sim 0.5\%$. In any case, if higher modeling accuracy on
$F^{\rm H}$ is needed, one can of course resort to a similar approach
to what Refs.~\cite{Yunes:2010zj,Nagar:2011aa} did for $F^{\infty}$,
namely fitting the numerical data. We pursue this task in
Appendix~\ref{sect:AppendixFitFH}.

\begin{figure}[!ht]
  \begin{center}
    \includegraphics*[width=0.45\textwidth]{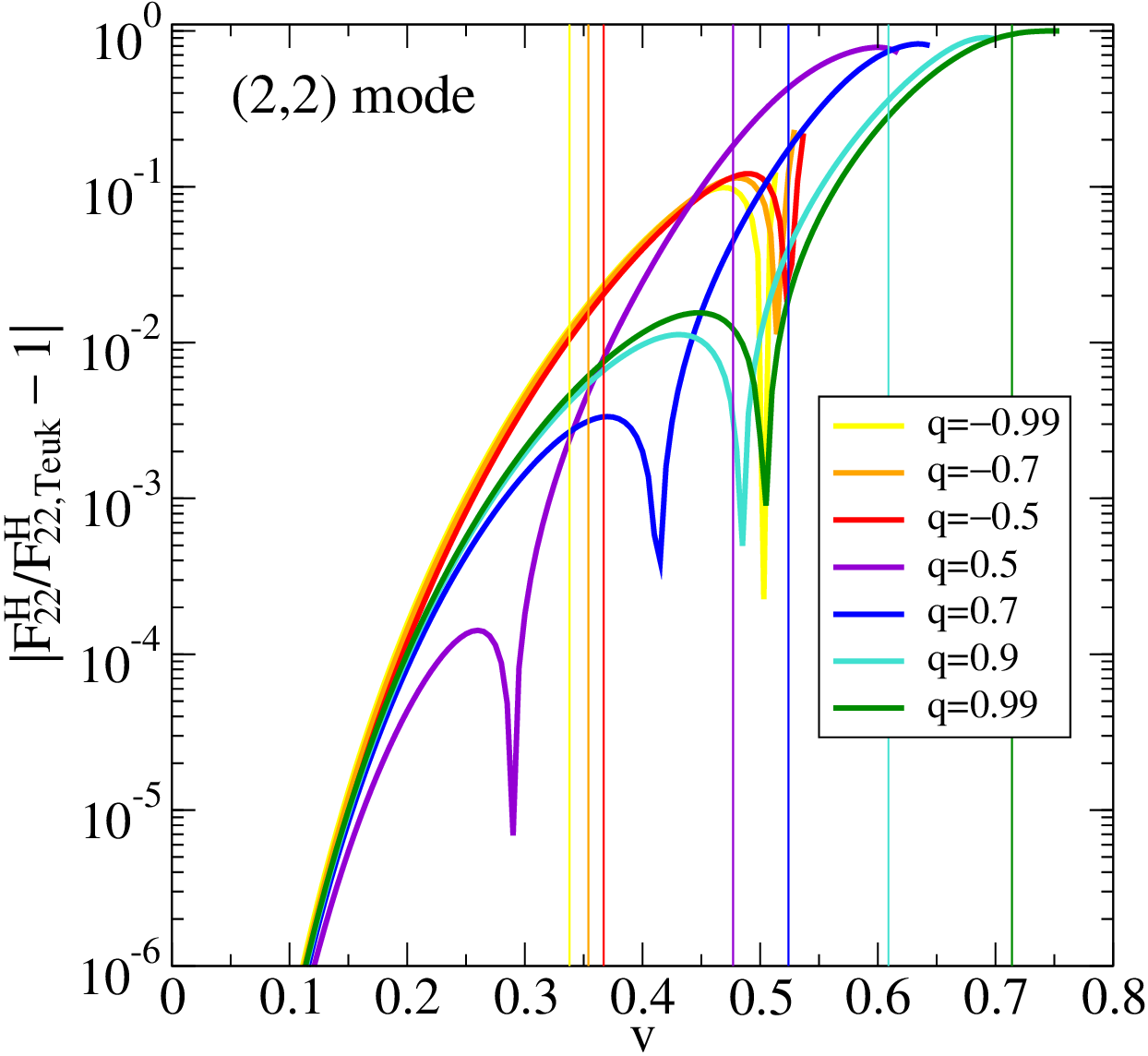}
    \caption{\label{fig:Err22} We show the fractional error of our model with respect to the (2,2) 
mode of the Teukolsky-equation BH-absorption flux. All curves extend up to the respective LRs. Vertical lines mark the
      positions of the respective ISCOs.}  
\end{center}
\end{figure}

\begin{figure}[!ht]
  \begin{center}
    \includegraphics*[width=0.45\textwidth]{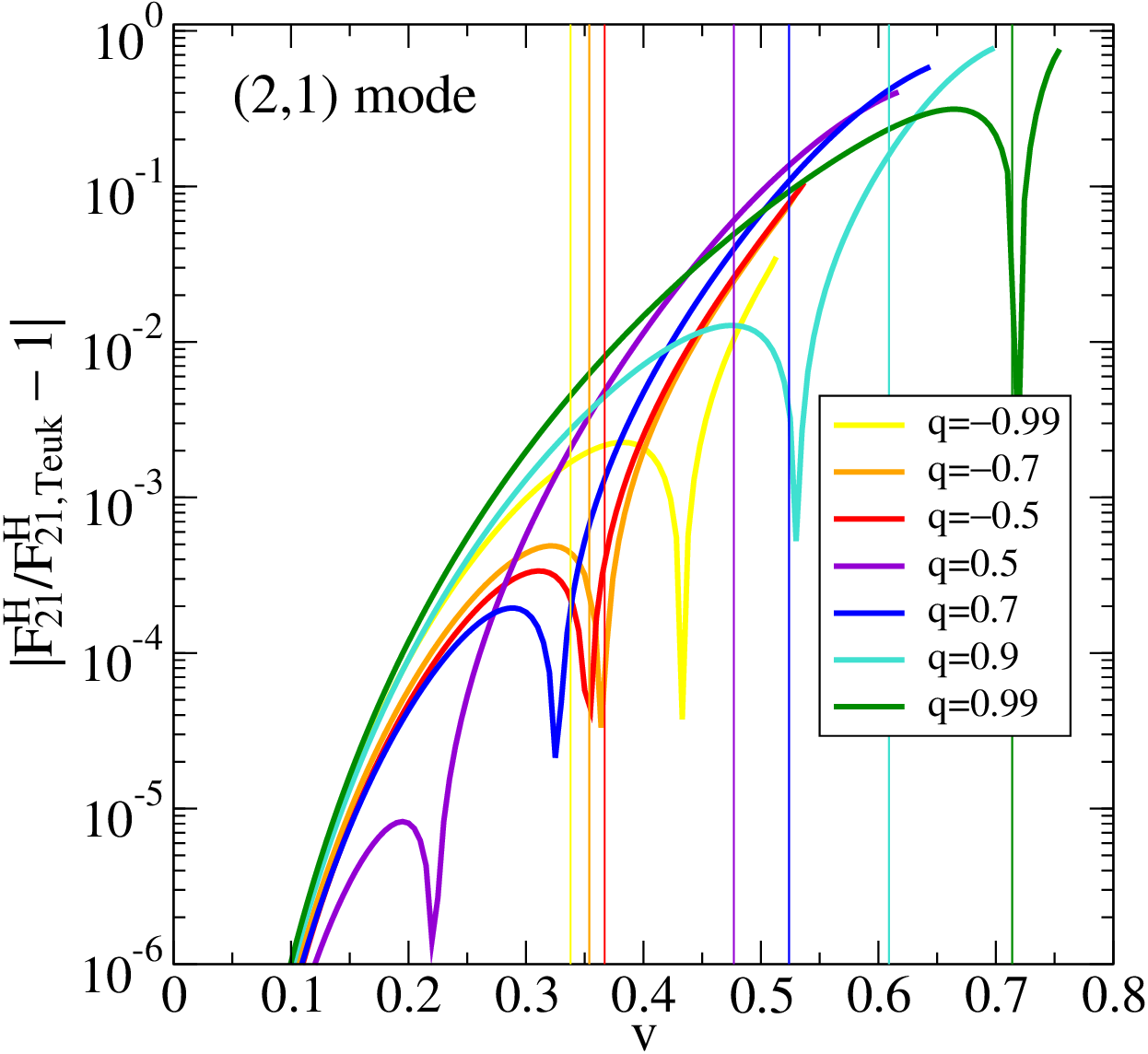}
    \caption{\label{fig:Err21} We show the fractional error of our model with respect to the (2,1) 
mode of the Teukolsky-equation BH-absorption flux. All curves extend up to the respective LRs. Vertical lines mark the
      positions of the respective ISCOs.}
  \end{center}
\end{figure}

Let us now focus on the multipolar modes of the BH-absorption flux,
rather than the total flux. In Figs.~\ref{fig:Err22} and
\ref{fig:Err21} we compare the dominant (2,2) mode and leading
subdominant (2,1) mode. We only show the results for the
$\tilde{\rho}^{\rm H}$--factorization, but comment also about the
performance of the Taylor flux below.  For the $\ell=2$ modes, the
relative error of our factorized model is at least one order of
magnitude smaller than the Taylor-expanded PN model across the entire
frequency range up to the LR. We also find that for the (3,3) mode the
improvement of the factorized model over the Taylor-expanded PN model
is more modest, especially at higher frequencies. For positive spins
the Taylor-expanded PN (3,3) mode has actually a comparable
performance to the factorized flux. This can be explained from the
fact that the analytical knowledge for $\ell=3,4$ modes is pretty
limited [see Eqs.~(I2)-(I7) in Ref.~\cite{Mino:1997bx}], so that the
two models cannot differ drastically.

\begin{figure}
  \begin{center}
   \includegraphics*[scale=0.4]{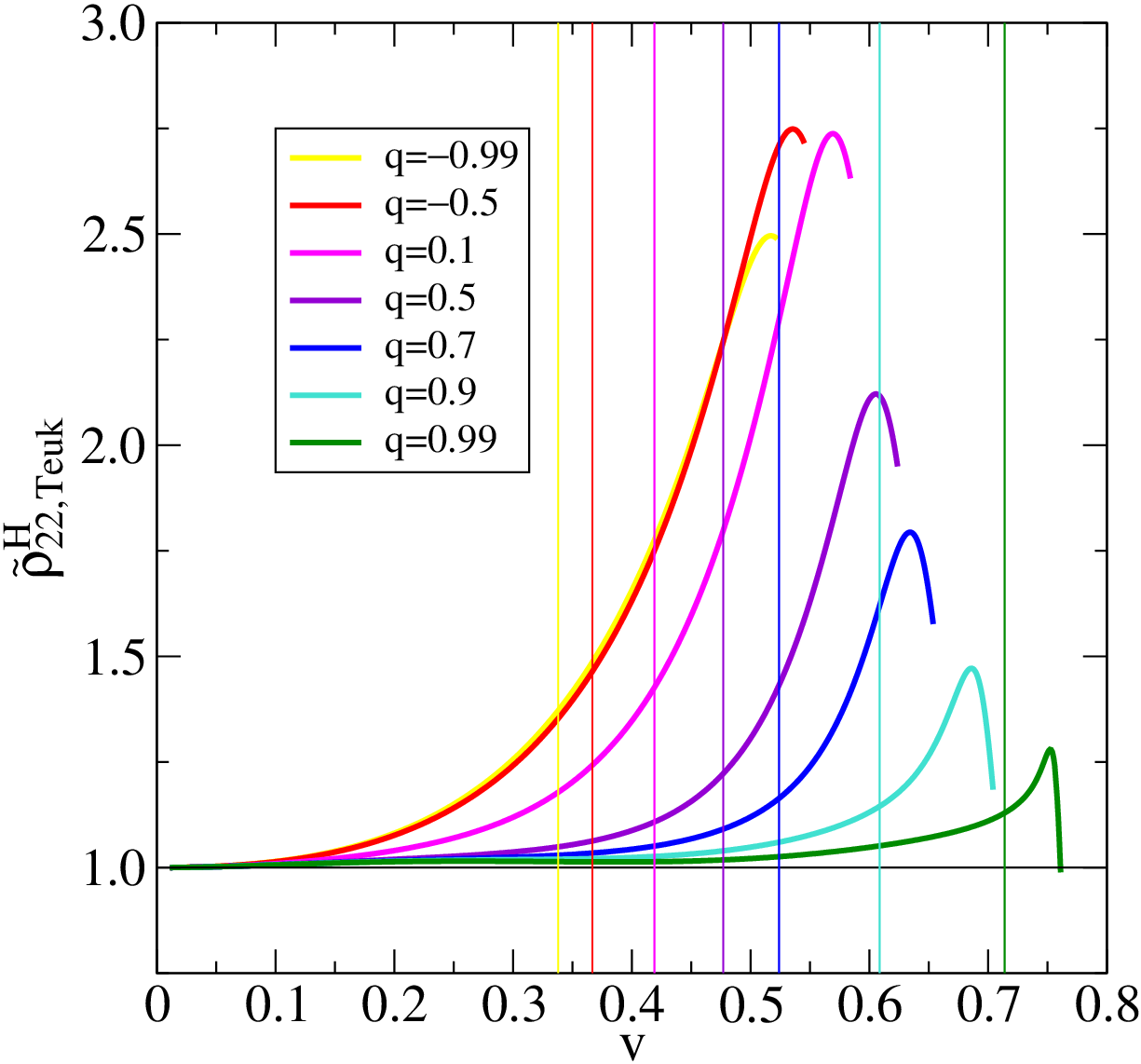}
  \caption{\label{fig:Teukrho} We show the Teukolsky-equation 
    $\tilde{\rho}^{\rm H}_{22}$ as functions of $v$. All 
    curves extend up to $r=r_{\rm LR}+0.01M$. As in the nonspinning case~\cite{Nagar:2011aa}, 
    also in the spinning case the Teukolsky-equation $\tilde{\rho}^{\rm H}_{22}$ behaves 
    non monotonically in the strong-field region close to the LS.  This peculiar behavior 
    cannot be easily captured by a polynomial model. This holds true also for 
    other modes with $\ell=m$. On the
    other hand, the Teukolsky-equation $\tilde{\rho}^{\rm H}_{\ell m}$'s for
    $\ell \neq |m|$ (e.g., the (2,1) mode) have monotonic 
    dependence on $v$ up to the LR. Vertical lines mark the positions
    of the respective ISCOs.}
\end{center}
\end{figure}

As we have discussed, in the factorized approach, the main ingredient
of modeling the absorption flux is the polynomial factor
$\tilde{\rho}_{\ell m}^{\rm H}$. Future progress in the PN knowledge
of the analytical fluxes will directly translate into new,
higher-order terms in the $\tilde{\rho}_{\ell m}^{\rm H}$
polynomials. Therefore it is useful to explicitly compute the
Teukolsky-equation $\tilde{\rho}_{\ell m,\rm Teuk}^{\rm H}$'s. We
simply divide $F^{\rm H}_{\ell m,\rm Teuk}$ by the leading and source
terms, and take the $2\ell$-th root. The result is shown in
Fig.~\ref{fig:Teukrho}, only for the $\ell=2$ modes. A peculiar
feature (generically seen in all modes with $\ell=m$) is the peak in
$\tilde{\rho}_{22,\rm Teuk}^{\rm H}$ in the strong-field regime,
inside the ISCO and close to the LR. Such feature is completely missed
by the polynomial model of Eq.~(\ref{rho22}).
Reference~\cite{Nagar:2011aa} noticed a similar shape in the
nonspinning limit, using their $\rho^{\rm H}_{22}$ mode [defined
  through Eq.~(\ref{BHAFactorization})], and proposed to fit it
through a rational function.  The $\tilde{\rho}_{21,\rm Teuk}^{\rm
  H}$'s do not display any relevant feature at high frequencies; this
is the case also for all the other $\ell \neq |m|$ modes that we
checked.  In Appendix~\ref{sect:AppendixFitFH} we provide a more
accurate analytical representation of the absorption flux by fitting
the Teukolsky-equation flux $F^{\rm H}$. These fits can be useful for
very accurate numerical evolution of PN or EOB equations of motions
for EMRIs, and also for the merger modeling of small mass-ratio binary
systems~\cite{TaracchiniSpinTPL}.

\subsection{Comparing black-hole absorption fluxes in the nonspinning case}

Before ending this section we want to compare our nonspinning results
to the numerical data and to the results of
Ref.~\cite{Nagar:2011aa}. As discussed above, the BH-absorption
Taylor-expanded PN flux is known through 6PN order beyond
$F^{\infty,\,\rm N}$ [see Eq.~(\ref{NSTaylorFlux})]. However, in
Ref.~\cite{Nagar:2011aa}, where the Schwarzschild case was considered,
the authors used the Taylor-expanded PN flux only through 5PN order
and, as a consequence, using Eq.~(\ref{BHAFactorization}) they
computed the BH-absorption factorized flux only up to 5PN order.
Using the full information contained in Refs.~\cite{Tagoshi:1997jy,
  Mino:1997bx} for the Taylor-expanded PN flux we obtain $\rho^{\rm
  H}_{22}$ through 6PN order, that is
\begin{equation}
\label{rhoH22}
\rho^{\rm H}_{22}(q=0) = 1+v^{2}+\frac{335}{84}v^{4}+\mathcal{O}(v^{6})\,.
\end{equation}
\begin{figure}[!ht]
  \begin{center}
    \includegraphics*[width=0.45\textwidth]{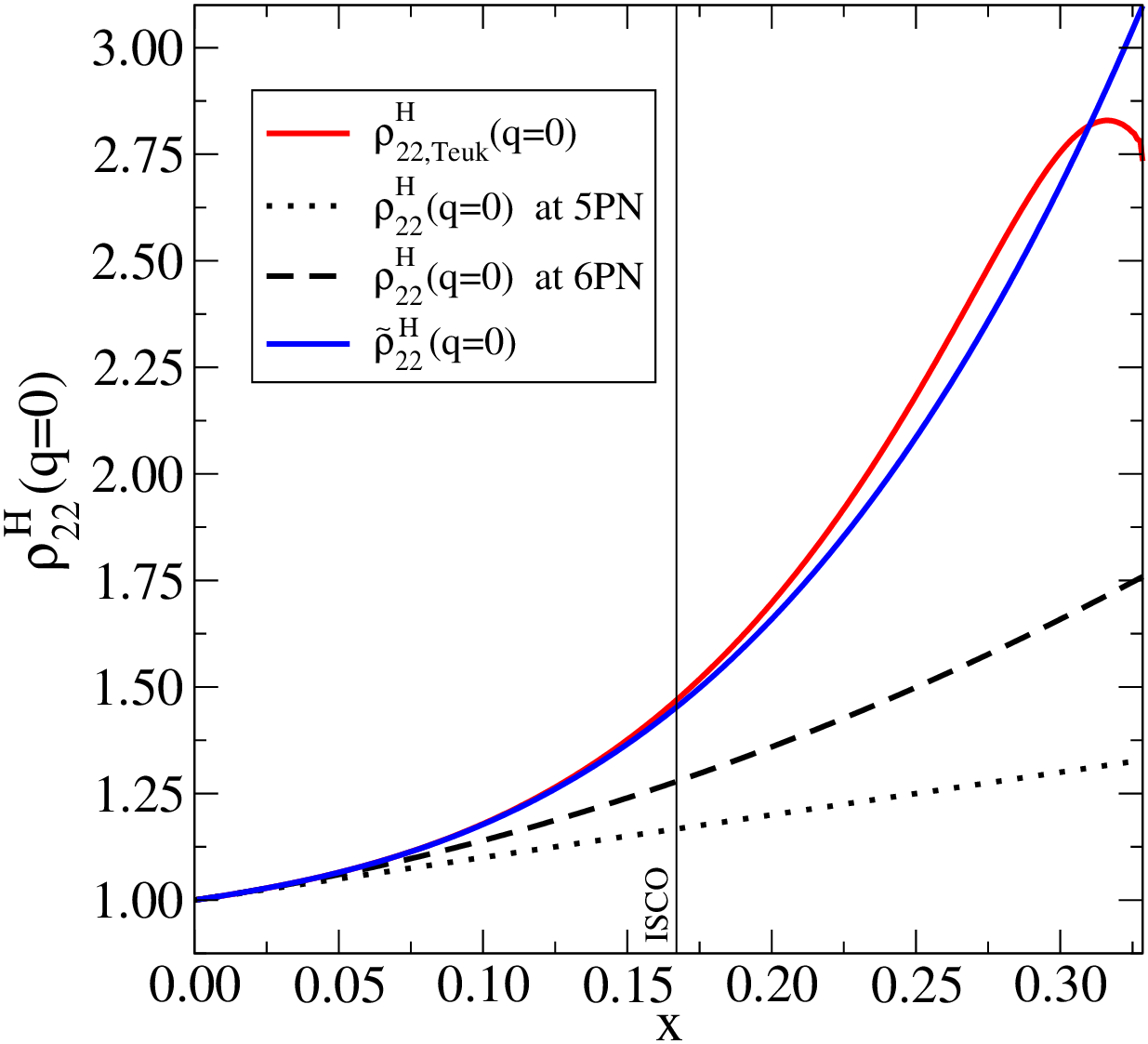}
    \caption{\label{fig:NSRho22} We compare the nonspinning
      $\rho_{22}^{\rm H}$ computed from the Teukolsky-equation data of
      $F_{22}^{\rm H}$ with the nonspinning factorized flux derived in 
      Ref.~\cite{Nagar:2011aa} up to 5PN order and in this paper up to
      6PN order. We also include the nonspinning limit of the factorized flux 
      $\tilde{\rho}_{22}^{\rm H}$ proposed in this paper. The curves are plotted against
      $x\equiv(M \Omega)^{2/3}=v^2$, and extend up to the LR in
      $x_{\rm LR}=1/3$. A vertical line marks the ISCO in $x_{\rm
        ISCO}=1/6$.}  \end{center}
\end{figure}
In Fig.~\ref{fig:NSRho22} we show for $q=0$ the $\rho^{\rm H}_{22}$ extracted from the numerical data as
\begin{equation}
\rho_{22,\,\rm Teuk}^{\rm H} (q=0) \equiv \left[\frac{2F_{22,\, \rm Teuk}^{\rm H}(q=0)}{\frac{32}{5}\left (\frac{\mu}{M}\right )^2 v^{18} (\hat{S}^{(0)}_{\rm eff})^2}\right]^{1/4}\,,
\end{equation}
the $\rho^{\rm H}_{22}$ at 5PN and 6PN order from Eq.~(\ref{rhoH22}), and the nonspinning limit of the $\tilde{\rho}_{22}^{\rm H}$ 
proposed in this paper.

It is interesting to observe that our $\tilde{\rho}_{22}^{\rm H}$ is much 
closer to the numerical data than the $\rho_{22}^{\rm H}$. We emphasize that 
in the nonspinning limit $\tilde{\rho}_{22}^{\rm H}$ contains higher-order PN terms produced by the factorization 
procedure, which singles out the zero $(1-\Omega/\Omega_H)$. 

\begin{figure}[!ht]
  \begin{center}
    \includegraphics*[width=0.45\textwidth]{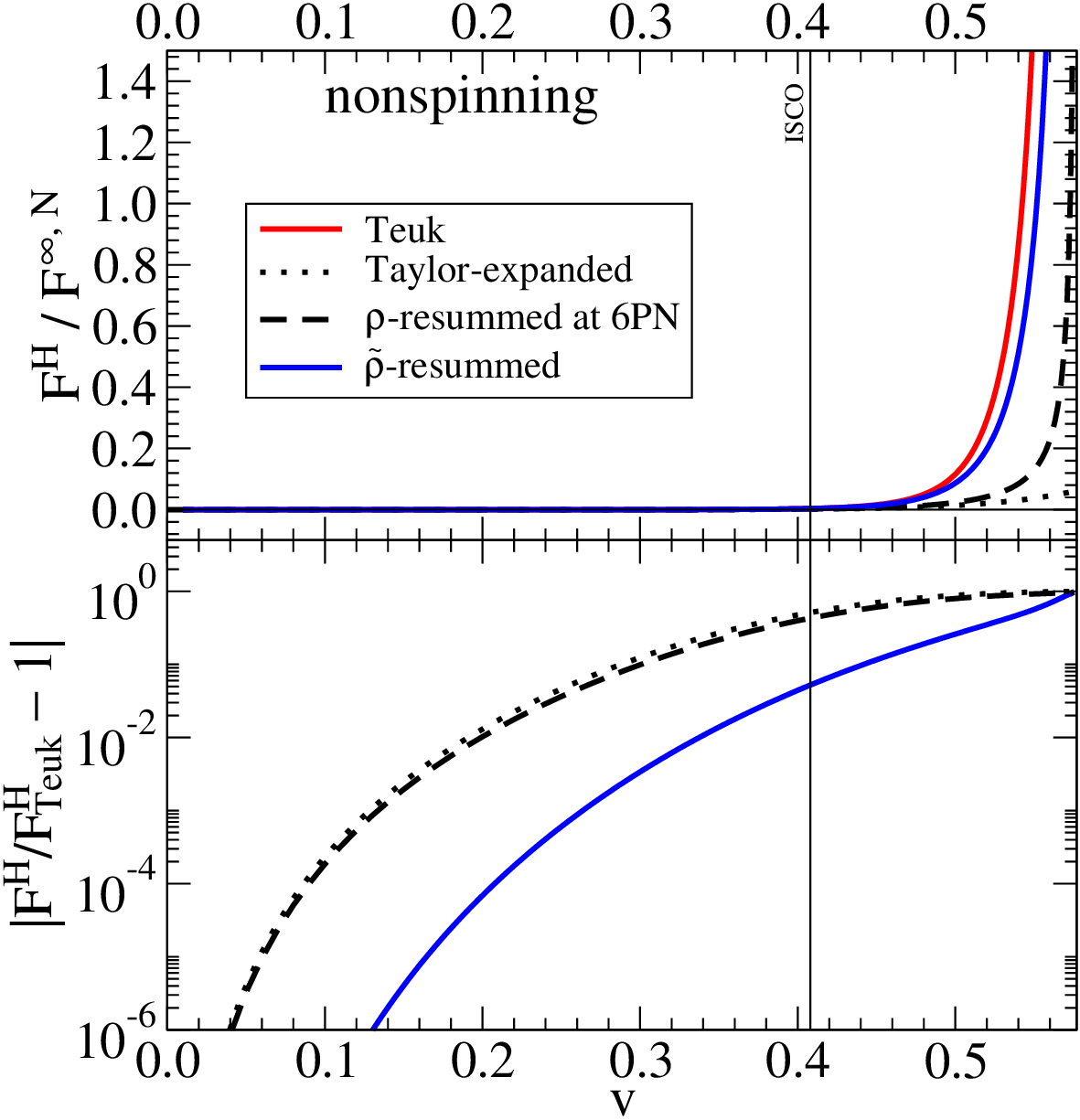}
    \caption{\label{fig:NSTot} We compare the nonspinning
        BH-absorption Teukolsky-equation flux to the nonspinning
        Taylor-expanded PN model of Ref.~\cite{Mino:1997bx} (see Eq.~(\ref{NSTaylorFlux})), the $\rho_{\ell
          m}^{\rm H}$--factorized model (see Eq.~(\ref{BHAFactorization})), and the nonspinning limit of the $\tilde{\rho}_{\ell
          m}^{\rm H}$--factorized model of this paper. A vertical line marks the ISCO. The curve extend up to
        the LR.}    
\end{center}
\end{figure}

For the sake of completeness, we list the rest of the $\rho_{\ell m}^{\rm H}$'s defined in Eq.~(\ref{BHAFactorization}) for $q=0$, which are computed starting from the nonspinning limit of the Taylor-expanded modes:
\begin{eqnarray}
\rho^{\rm H}_{21}(q=0) &=& 1+\frac{19}{12}v^{2}+\mathcal{O}(v^{4})\,,\\
\rho^{\rm H}_{33}(q=0) &=& \rho^{\rm H}_{31}(q=0)=1+\mathcal{O}(v^{2})\,,\phantom{\frac{19}{12}}\\
\rho^{\rm H}_{32}(q=0) &=& \rho^{\rm H}_{4m}(q=0)=\mathcal{O}(v)\,.\phantom{\frac{19}{12}}
\end{eqnarray}

Lastly, in Fig.~\ref{fig:NSTot} we consider the nonspinning limit and compare the BH-absorption 
total numerical flux to the nonspinning (i) Taylor-expanded PN  
flux~\cite{Mino:1997bx}, (ii) the $\rho^{\rm H}_{\ell m}$ factorized flux from 
Eq.~(\ref{NSTaylorFlux}) and (iii) the $\tilde{\rho}^{\rm H}_{\ell m}$ factorized flux proposed in this paper 
and given in Eq.~(\ref{BHAFactorizationTilde}). 

\section{Conclusions}
\label{sec:Conclusions}

Building on Refs.~\cite{DIN, Pan2010hz,Nagar:2011aa}, we have proposed
a new analytical model for the BH-absorption energy flux of a test
particle on a circular orbit in the equatorial plane of a Kerr BH.  We
recast the Taylor-expanded PN flux in a factorized form that allowed
us to enforce two key features present in BH perturbation theory: the
presence of a zero at a frequency equal to the frequency of the
horizon, and the divergence at the LR. The latter was also adopted for
the energy flux at infinity in
Refs.~\cite{DIN,Pan2010hz,Nagar:2011aa}.  These features are not
captured by the Taylor-expanded PN flux. We compared our model to the
absorption flux computed from the numerical solution of the Teukolsky
equation in frequency
domain~\cite{Hughes:1999bq,Hughes:2001jr,Drasco:2005kz}. In
particular, we computed the gravitational-wave fluxes both at infinity
and through the horizon for a Kerr spin $-0.99 \leq q \leq 0.99$, and
for the first time down to a radial separation $r = r_{\rm
  LR}+0.01M$. This extended previous work~\cite{Yunes:2010zj} to
unstable circular orbits below the ISCO.

We investigated the hierarchy of the multipolar flux modes. As the spin grows to large positive values,
more and more modes become comparable to the dominant (2,2) mode, even
before the ISCO. Among modes with the same value of $\ell$, the dominant ones are those with $\ell=|m|$. Close to the LR all modes with $\ell=|m|$ become
comparable to the (2,2) mode. We also studied how the mode hierarchy changes at the ISCO frequency when we vary the spin. We found that
only the $(2,1)$ and $(3,3)$ modes are always larger than the $(2,2)$
mode by more than $1\%$; only when $q \gtrsim 0.95$ the (4,4), (5,5),
(6,6) and (3,2) modes are above the $1\%$ threshold at the ISCO. One can understand these facts analytically within the WKB approximation, as already pointed out by old studies on geodesic synchrotron radiation~\cite{Davis:1972dm, Breuer:1973kt, Chrzanowski:1974nr,BreuerBook}. One can rewrite the radial Teukolsky equation in a Schr\"{o}dinger-like form, so that the flux modes turn out to be proportional to a barrier-penetration factor which exponentially suppresses modes with $\ell \neq |m|$. 

We compared the numerical fluxes at infinity and through the horizon
with the factorized fluxes for several spin values $-0.99 \leq q\leq
0.99$. For the energy flux at infinity, we found that the factorized
model developed in Ref.~\cite{Pan2010hz} is reliable for retrograde
orbits and in the nonspinning case almost up to the LR, but performs
rather poorly for large spin prograde orbits close to the LR. For the
BH-absorption energy flux we found that the agreement of the
factorized flux to the numerical flux is always better than the one of
the Taylor-expanded PN flux. The fractional difference between the
numerical and factorized flux is less than $1\%$ up to the ISCO for
$-1 \le q \leq 0.5$. For spins $q>0.7$ the factorized flux starts
performing worse, even before the ISCO, but it always performs better
than the Taylor-expanded PN flux. We expect that the large modeling
error after the ISCO for $q>0.7$ will not affect much the inspiral,
merger and ringdown waveforms produced with the time-domain Teukolsky
equation evolved with the factorized flux. In fact, the energy flux
does not have much effect beyond the ISCO, since the system's dynamics
at that point are well described by a plunging geodesic. In
Ref.~\cite{TaracchiniSpinTPL} we show that evolving an EOB dynamics
with the factorized model instead of the numerical flux introduces a
difference in the time of coalescence smaller than half of a percent
across the whole spin range.

Finally, in the Appendices~\ref{sect:AppendixFitFInf} 
and \ref{sect:AppendixFitFH} we computed fits to the numerical 
fluxes at infinity and through the horizon that could be used 
for highly accurate numerical evolution of EMRIs using PN or EOB equations of motions, 
and also for modeling the merger waveforms of small mass-ratio binary systems~\cite{TaracchiniSpinTPL}.

Future work may address the issue of why the total energy fluxes
normalized to the specific energy diverge when computed exactly at the
photon orbit.  In fact, as we discussed, in this case the WKB
treatment suggests a non-convergent sum over the multipolar modes
[see Eq.~(\ref{Fell})], since the cutoff mode index $\ell_{c}$
would go to $+\infty$.

In the near future we plan to extend the factorized model of the
BH-absorption flux to the case of spinning comparable-mass BHs, so
that it can be used in the EOB model when calibrating it to
numerical-relativity simulations.

\begin{acknowledgments}

We thank Enrico Barausse, Yi Pan and Nico Yunes for useful and informative discussions. 

A.B. and A.T. acknowledge partial support from NSF Grants
No. PHY-0903631 and No. PHY-1208881.  A.B. also acknowledges partial
support from the NASA Grant NNX09AI81G and A.T. from the Maryland
Center for Fundamental Physics. A.B. and A.T. also thank the
hospitality of the Kavli Institute for Theoretical Physics (supported by
the NSF Grant No. PHY11-25915) where part of this work was carried
out. This work was supported at MIT by NSF Grant PHY-1068720.  SAH
gratefully acknowledges fellowship support by the John Simon
Guggenheim Memorial Foundation, and sabbatical support from the
Canadian Institute for Theoretical Astrophysics and the Perimeter
Institute for Theoretical Physics. G.K. acknowledges research 
support from NSF Grant Nos. PHY-1016906, CNS-0959382 and PHY-1135664,  
and from the US Air Force Grant Nos. FA9550-10-1-0354 and 10-RI-CRADA-09.
\end{acknowledgments}

\appendix

\section{The Teukolsky-equation source term for light-ring orbits}
\label{app:LRsource}

In this appendix, we describe how the divergence in fluxes at the
light ring enters through the Teukolsky equation's source term, as
well as a simple modification that allows us to factor it from the
flux computation.  This divergence-free form proved useful for
understanding how fluxes behave in the extreme strong field.

We begin with the stress-energy tensor of a body with rest mass $\mu$
moving in the Kerr spacetime:
\begin{equation}
T_{\alpha\beta} = \mu \int u_\alpha u_\beta\, \delta^{(4)}[x^\mu -
  z^\mu(\tau)]\,d\tau\;.
\label{eq:particlesource}
\end{equation}
Here, $x^\mu$ is a general spacetime coordinate, and $z^\mu(\tau)$ is
the worldline followed by the moving body; $u^\alpha =
dz^\alpha/d\tau$, where $\tau$ is proper time along the worldline.
The delta function is normalized so that
\begin{equation}
\int \sqrt{-g}\,\delta^{(4)}\,d^4x = 1\;,
\end{equation}
where $g = -\Sigma \sin^2\theta$ is the determinant of the Kerr
metric, and $\Sigma = r^2 + q^2M^2\cos^2\theta$.

In a typical particle analysis, we integrate
Eq.\ (\ref{eq:particlesource}) immediately to obtain
\begin{equation}
T_{\alpha\beta} = \mu \frac{u_\alpha u_\beta}{\Sigma\sin\theta (dt/d\tau)}\,
\delta[r - r(t)]\,\delta[\theta - \theta(t)]\,\delta[\phi - \phi(t)]\;.
\label{eq:particlesource2}
\end{equation}
This is well-behaved except when $dt/d\tau \to 0$.  This occurs at the
light ring, and explains why gravitational-wave fluxes diverge as the
light ring is approached.

Let us rewrite Eq.\ (\ref{eq:particlesource}) using $d\lambda =
d\tau/\mu$, in anticipation of taking the limit $\mu \to 0$.  Using
the fact that $dz^\alpha/d\lambda = p^\alpha$, the momentum of the
body, we find
\begin{eqnarray}
T_{\alpha\beta} &=& \frac{1}{\mu} \int p_\alpha p_\beta\,
\delta^{(4)}[x^\mu - z^\mu(\lambda)](\mu\,d\lambda)
\nonumber\\
&=& \int p_\alpha p_\beta\, \delta^{(4)}[x^\mu - z^\mu(\lambda)]\,d\lambda\;.
\label{eq:particlesource3}
\end{eqnarray}
This is easily integrated, and we find
\begin{eqnarray}
T_{\alpha\beta} &=& \frac{p_\alpha p_\beta}{\Sigma\sin\theta\,p^t}\,
\delta[r - r(t)]\,\delta[\theta - \theta(t)]\,\delta[\phi - \phi(t)]
\nonumber\\
&=& \frac{p_\alpha p_\beta}{\Sigma\,p^t}\,
\delta[r - r_{\rm o}]\,\delta[\theta - \pi/2]\,\delta[\phi - \phi(t)]\;.
\label{eq:particlesource4}
\end{eqnarray}
On the second line, we specialize to a circular orbit of radius $r =
r_{\rm o}$ in the equatorial plane.  Equation
(\ref{eq:particlesource4}) is well behaved as $\mu \to 0$.

The momenta which appear in this stress-energy tensor are determined
by the geodesic equations for Kerr orbits~\cite{Bardeen:1972fi}
\begin{eqnarray}
\Sigma\,p^t &=& \frac{(r^2 + q^2M^2)}{\Delta}\left[E(r^2 + q^2M^2) - qM
  L_z\right]
\nonumber\\
& & + q M (L_z - q M E)\;,
\label{eq:pt}\\
\Sigma\,p^\phi &=& \frac{qM}{\Delta}\left[E(r^2 + q^2 M^2) - q M L_z\right]
 + L_z - q M E\;,
\nonumber\\
\label{eq:pphi}\\
\left(\Sigma\,p^r\right)^2 &=& \left[E(r^2 + q^2 M^2) - q M L_z\right]^2
\nonumber\\
& & - \Delta\left[\mu^{2} r^2 + \left(L_z - q M E\right)^2\right]\;.
\label{eq:pr}
\end{eqnarray}
We have specialized to $\theta = \pi/2$.  This allows us to set the
Carter constant $Q = 0$ and to neglect $p^\theta$.

Equations (\ref{eq:pt}) and (\ref{eq:pphi}) are proportional to the
orbiting body's rest mass $\mu$; Eq.\ (\ref{eq:pr}) is proportional to
$\mu^2$.  In most Teukolsky solvers, we factor out the overall factors
of $\mu$, and thereby express everything on a per-unit-rest-mass
basis.  As the light ring is approached, the energy and angular
momentum per unit rest mass diverge.  In anticipation of this, let us
instead divide by the orbital energy $E$.  Defining $\hat p^\mu \equiv
p^\mu/E$, the stress-energy tensor is written
\begin{eqnarray}
T_{\alpha\beta} &=& E\,\frac{\hat p_\alpha \hat
  p_\beta}{\Sigma\sin\theta\,\hat p^t}\, \delta[r -
  r(t)]\,\delta[\theta - \theta(t)]\,\delta[\phi - \phi(t)]
\nonumber\\
&=& E\,\frac{\hat p_\alpha\hat p_\beta}{\Sigma\,\hat p^t}\, \delta[r -
  r_0]\,\delta[\theta - \pi/2]\,\delta[\phi - \phi(t)]\;,
\label{eq:particlesource5}
\end{eqnarray}
where again the second line is specialized to an equatorial, circular
orbit.  The momenta appearing here are given by
\begin{eqnarray}
\Sigma\, \hat p^t &=& \frac{(r^2 + q^2M^2)}{\Delta}\left[(r^2 + q^2M^2)
  - q M b\right]
\nonumber\\
& & + q M(b - q M)\;,
\label{eq:hatpt}\\
\Sigma\, \hat p^\phi &=& \frac{qM}{\Delta}\left[(r^2 + q^2M^2) - q M b\right]
 + b - qM\;,
\nonumber\\
\label{eq:hatpphi}\\
\left(\Sigma\, \hat p^r\right)^2 &=& \left[(r^2 + q^2M^2) - q M b\right]^2
\nonumber\\
& &-\Delta\left[\frac{r^2}{\hat{E}^{2}} + \left(b - q M\right)^2\right]\;.
\label{eq:hatpr}
\end{eqnarray}
We have introduced the orbit's energy per unit rest mass $\hat E\equiv
E/\mu$ and the orbit's ``impact parameter'' $b \equiv L_{z}/E$ [see
  Eqs.~(\ref{Ecirc})--(\ref{Lcirc})].  These expressions work well all
the way to the light ring, Eq.\ (\ref{eq:lightring}).

To implement this form of the source, we follow the recipe outlined in
Sec.\ IV of Ref.\ {\cite{Hughes:1999bq}} [see especially Eqs.\ (4.32)
  -- (4.34)], but using Eq.\ (\ref{eq:particlesource5}) instead of
Eq.\ (\ref{eq:particlesource2}).  The code then computes the
amplitudes $Z^\star_{\ell m}$ per unit orbital energy rather than per
unit rest mass, and hence computes all fluxes per unit orbital energy
squared.  This factors out the divergence associated with the behavior
of the energy per unit mass at the light ring.

When this is done, each modal contribution $F^\star_{\ell m}$ is
perfectly well behaved at the light ring.  The sum of all modes can
grow quite large, but only because there are many modes that
contribute, not because of the pole at the light ring.

\section{Expressions for $\tilde{f}^{\rm H}_{\ell m}$}
\label{sect:Appendixflm}

In this appendix we write the explicit expressions of the
$\tilde{f}^{\rm H}_{\ell m}$ polynomials. We find
\begin{widetext}
\begin{subequations}
\label{f2}
\begin{align}
 \tilde{f}^{\rm H}_{22}&=1 + 2 v^2 - \bigg\{4 B_2 + \frac{2q}{\kappa\left(1+3q^2\right)} \left[5 + 4\kappa-q^2 \left(2  + 3 q^2\right)\right]\bigg\}v^3 +  \left(\frac{377}{42} -\frac{8}{42} q^2\right) v^4 \notag \\
&-\bigg\{8B_2+\frac{q}{1+3q^2} \left[ \frac{119}{9}-\frac{25}{3}q^2+4\kappa\left(5+3q^2\right)\right]\bigg\} v^5 + \bigg\{\frac{547\,402}{11\,025} - \frac{4}{3}\pi^2 -\frac{7\,942}{567}q^2+2q^4+8B_2^2\notag\\
&+8C_2 \bigg(1+\frac{2}{\kappa}\bigg) -\frac{856}{105}\left(A_2+\gamma_E+\log{2}+\log{\kappa}+2\log{v}\right)-\frac{1}{1+3q^2}\left[\frac{152}{9}-32 q B_2-8q \kappa B_2\left(5+3q^2\right)\right]\notag\\
&+\frac{1}{\left(1+3q^2\right)^2}\left[\frac{224}{9}+4\kappa\left(5+4q^2+9q^4-18q^6\right)\right]\bigg\} v^6-\bigg[-\frac{1\,641}{189}q+\frac{73}{189}q^3+\frac{4\,556q}{63\left(1+3q^2\right)}\notag\\
&+\frac{1}{21}\left(377-8q^2\right)\left(2B_2+q\kappa \frac{5+3q^2}{1+3q^2}\right)\bigg]v^7+ \bigg\{\frac{4\,579\,699}{33\,075} - \frac{8}{3} \pi^2 -\frac{14\,617}{567}q^2+\frac{529}{126}q^4-\frac{5\,296}{105}\gamma_E-\frac{1\,712}{105}A_2\notag\\
&+16B_2^2+16C_2\left(1+\frac{2}{\kappa}\right)-\frac{100}{9}qB_2 + \frac{1}{1+3q^2}\left[-\frac{712}{27}+64qB_2+16\kappa q B_2\left(5+3q^2\right)\right]+ \frac{1}{\left(1+3q^2\right)^2}\notag\\
&\times\bigg[\frac{448}{9}+\kappa \bigg(40 + \frac{38}{9} q^2 - 28 q^4  
-194 q^6\bigg)\bigg]-\frac{592}{7}\log{2}-\frac{1\,712}{105}\log{\kappa}-\frac{2\,336}{35}\log{v}\bigg\} v^8+ \mathcal{O}(v^9)\,, \label{f22}\\
\tilde{f}^{\rm H}_{21}&=1-\frac{2}{3}q v+\frac{7}{6}v^2+ \bigg\{-2B_1 +\frac{2q}{4 - 3 q^2}\left[\frac{5}{3}-2q^2-\kappa\left(5-3q^2\right)\right]\bigg\}v^3+\bigg\{\frac{841}{504}+\frac{4}{3}q B_1\notag\\
&-\frac{1\,165}{378}q^2 +\frac{4}{3\left(4-3q^2\right)}\bigg[\frac{4}{3}+q^2\kappa\left(5-3q^2\right)\bigg]\bigg\}v^4+\bigg\{\frac{785}{252}q+\frac{13}{14}q^3-\frac{7}{3}\left[B_1+\frac{q}{4-3q^2}\left[1+\kappa\left(5-3q^2\right)\right]\right]\bigg\}v^5\notag\\
& + \bigg\{\frac{303\, 727}{19\, 600}-\frac{12\, 055}{2\, 268}q^2+2q^4-\frac{\pi^2}{3}-\frac{214}{105}\left(A_1+\gamma_E+\log{2}+\log{\kappa}+2\log{v}\right)+2 B_1^2\notag\\
&+2C_1\bigg(1+\frac{2}{\kappa}\bigg)-\frac{1}{4-3q^2} \bigg[\frac{40}{9}
-4q B_1\left(-\frac{5}{3}+2q^2+\kappa\left(5-3q^2\right)\right)\bigg] +\frac{16}{3\left(4-3q^2\right)^2}\bigg[\kappa\left(15-52q^2\right.\notag\\
&\left.+54q^4 -18q^6\right)-\frac{1}{3}\bigg]\bigg\}v^6+ \mathcal{O}(v^7)\label{f21} \,,
\end{align}
\end{subequations}
\begin{subequations}
\label{f3}
\begin{align}
\tilde{f}^{\rm H}_{33}&=1+\frac{7}{2}v^2 -\bigg\{6B_3+\frac{q}{\left(1+8q^2\right)\left(4+5q^2\right)}\bigg[\frac{262}{3}+\frac{628}{3}q^2-\frac{80}{3}q^4 + 18\kappa\left(5+13q^2\right)\bigg] \bigg\} v^3\notag\\
&+\bigg(\frac{1\,549}{120} - \frac{5}{6} q^2\bigg)  v^4+\mathcal{O}(v^5)\,,\label{f33}\\
\tilde{f}^{\rm H}_{32}&=1-\frac{3}{4}q v+\frac{5}{2} v^2+\mathcal{O}(v^3)\label{f32} \,,\\
\tilde{f}^{\rm H}_{31}&=1+ \frac{29}{6}v^2 -2\bigg\{B_1 + \frac{q}{4-3q^2}\bigg[\kappa \left(5-3q^2\right)+\frac{1}{9-8q^2}\bigg(65-\frac{866}{9}q^2+\frac{104}{3}q^4\bigg)\bigg]\bigg\}v^3 \notag\\
&+\bigg(\frac{1\,195}{72}+\frac{1}{2}q^2\bigg)v^4+\mathcal{O}(v^5)\label{f31} \,,
\end{align}
\end{subequations}
\begin{subequations}
\label{f4}
\begin{align}
\tilde{f}^{\rm H}_{44}&=1+\mathcal{O}(v)\,,\label{f44}\\
\tilde{f}^{\rm H}_{43}&=\mathcal{O}(v)\label{f43} \,,\\
\tilde{f}^{\rm H}_{42}&= 1+\mathcal{O}(v)\label{f42} \,,\\
\tilde{f}^{\rm H}_{41}&=\mathcal{O}(v)\label{f41} \,.
\end{align}
\end{subequations}
\end{widetext}
We have compared the factorized fluxes built using either the $\tilde{\rho}^{\rm H}_{\ell m}$'s or  $\tilde{f}^{\rm H}_{\ell
  m}$'s against the Teukolsky-equation flux and have found that the latter have fractional differences one order of 
magnitude smaller than the former for prograde orbital geometries. For retrograde orbits, instead, the two factorizations have more similar modeling errors. For this reason we have employed the 
 $\tilde{\rho}^{\rm H}_{\ell m}$--factorization in the paper.

\section{Fits of the gravitational flux at infinity}
\label{sect:AppendixFitFInf}

In this appendix we fit the gravitational flux at infinity computed through the 
Teukolsky equation to further improve the amplitude of the factorized modes, given in 
Eq.~(\ref{InfinityFactorization}), and the total factorized flux. 
The Teukolsky-equation data available to us span frequencies from $v=0.01$ up to $r=r_{\rm
  LR}+0.01M$, and have spins in the range $q\in\{-0.99$, $-0.95$,
$-0.9$, $-0.8$, $-0.7$, $-0.6$, $-0.5$, $-0.4$, $-0.3$, $-0.2$,
$-0.1$, 0, 0.1, 0.2, 0.3, 0.4, 0.5, 0.6, 0.7, 0.8, 0.9, 0.95,
0.99$\}$.

Improving the mode's amplitude $|h_{l m}|$'s (which is equivalent to
improving the mode's flux $F^{\infty}_{l m}$) is conducive to the EOB
modeling of the merger signal in the small mass-ratio limit for large
spins, which we have pursued in Ref.~\cite{TaracchiniSpinTPL}. [Note
  that the modes in this appendix are spherical-harmonic modes,
  labeled $(l,m)$.] Earlier efforts in this direction (e.g., see
Refs.~\cite{Yunes:2010zj,Barausse:2011kb}) were plagued by significant
modeling errors in the $|h_{l m}|$'s for spins $q\gtrsim 0.7$. For
such systems, the discrepancies between time-domain Teukolsky-equation
waveforms and EOB waveforms showed up early on during the adiabatic
inspiral, where non-quasi-circular effects are still negligible. This
had also the effect of introducing a large error on the total
$F^{\infty}$, which depends on the $|h_{l m}|$'s through
Eq.~(\ref{FInfModes}).

We perform the fit by adding to the $\rho_{l m}$'s of Ref.~\cite{Pan2010hz} 
an additional term $\rho_{l m}^{\rm  amp\,fit}$, which is determined by the fit. 
We fit the minimal number of unknown higher PN orders beyond the current analytical
knowledge of the $\rho_{l m}$'s, such that the residuals on the
individual $F^{\infty}_{l m}$ (or, equivalently, on $|h_{l m}|$)
are within $5\%$ up to the ISCO. It is worth reminding that
Ref.~\cite{Pan2010hz} based their factorized model on unpublished
Taylor-expanded modes computed in BH perturbation theory by Tagoshi
and Fujita. In previous years, Ref.~\cite{Tanaka:1997dj} had derived
the Taylor-expanded modes needed to compute the 5.5PN energy flux at
infinity for the Schwarzschild case, while Ref.~\cite{Tagoshi:1996gh}
had derived the Taylor-expanded modes needed to compute the 4PN energy
flux at infinity for a particle in the equatorial plane of a Kerr
BH. However, in both instances, the explicit formulae had not been
published. Reference~\cite{Fujita:2010xj} independently derived the
nonspinning Taylor-expanded multipolar waveforms up to 5.5PN order,
and computed a $\rho_{l m}-$factorization which includes some
higher PN nonspinning terms as compared to
Ref.~\cite{Pan2010hz}. Reference~\cite{Pan2010hz} itself pointed out
(before Eq.~(A1)) that their nonspinning $\rho_{l m}$'s agreed with
those of Ref.~\cite{Fujita:2010xj} only up to ${\cal
  O}(v^{11-2(l-2)})$. References~\cite{Fujita:2011zk} and \cite{Fujita:2012cm} pushed the computation of the energy flux at infinity for Schwarzschild up to 14PN and 22PN order respectively, but provided only the 6PN term entering the $\rho_{22}$. Again, for the rest of this appendix we will
build upon the analytical results of Ref.~\cite{Pan2010hz}.

Table~I of Ref.~\cite{Pan2010hz} lists the PN knowledge of the
different modes $h_{l m}$'s at the time of publication. In
particular, given $(l,m)$, from the second line of that table one
can read the available PN order beyond the leading term $h_{l
  m}^{(N,\epsilon)}$ for the Taylor-expanded expression of the mode,
with a distinction between nonspinning and spinning terms. It turns
out that when $l \leq 5$ the nonspinning sector is known to a
higher or equal PN order than the spinning sector; on the other hand,
when $l > 5$ the knowledge of the spinning terms is better than the
nonspinning ones.

As already pointed out in
Refs.~\cite{Finn:2000sy,Pan2010hz,Barausse:2011kb,Barausse:2011vx},
the larger the value of $q$, the more multipolar modes become
comparable with the dominant (2,2) mode: see Fig.~3 of
Ref.~\cite{Barausse:2011kb}, which shows the mode hierarchy for
$q=0,0.9$ based on their amplitude $|h_{l m}|$.  An analytical
explanation for the hierarchy of the modes can also be found using the
WKB approximation~\cite{Davis:1972dm, Breuer:1973kt,
  Chrzanowski:1974nr,BreuerBook}. The multipolar modes we fit are:
(2,2), (2,1), (3,3), (3,2), (3,1), (4,4), (4,3), (4,2), (5,5), (5,4),
(6,6), (7,7) and (8,8).

Note that we perform the fits in the domain of the orbital velocity
$v\equiv (M \Omega)^{1/3}$, over the restricted range $0.01 \leq v
\leq v_{\rm ISCO}$ (where $v_{\rm ISCO}\equiv (M\Omega_{\rm
  ISCO})^{1/3}=[(r_{\rm ISCO}/M)^{3/2}+q]^{-1/3}$). The reason for
doing so (instead of going up to the final available frequency) is
threefold: i) from the point of view of the waveform, our primary goal
is to improve the adiabatic analytical model and modeling errors in
the plunge amplitude can easily be fixed by introducing
non-quasi-circular corrections~\cite{TaracchiniSpinTPL}; ii) from the
point of view of the energy flux at infinity, after the ISCO the
orbital motion of the binary becomes basically geodetic!\footnote{The
  plunge lasts for a time $\mathcal{O}(M)$, in contrast to the
  inspiral, which lasts for a much longer time
  $\mathcal{O}(M^{2}/\mu)$~\cite{Buonanno00,Mino:2008at}. Therefore
  the motion of the plunging particle is well approximated by a
  geodesic in Kerr spacetime.}; iii) we find it difficult to fit well 
  the post-ISCO data, all the way to the LR without spoiling the 
  low-frequency portion of the fit. As to the spin
range covered, we cannot include $q=0.99$ without affecting in a
negative way smaller spins. While computing the fits, we give equal
weight to all available spins, and fit them all together. This is
achieved by rescaling each range $0.01\leq v \leq v_{\rm ISCO}$
such that they all have the same measure, and by stitching together all
different ranges. We also tried fits in the domain of the orbital
frequency $M \Omega$, which amounts to giving more importance to
higher frequencies, but this created large relative errors at lower
frequencies, where the binary spends the majority of the time, therefore 
increasing the phase error due to the flux modeling.

Table~\ref{tab:Fits} lists the fitted functions $\rho_{l m}^{\rm
  amp\,fit}$. In those expressions we use $\textrm{eulerlog}_m x
\equiv \log{\gamma_E} + \log{2m}+\log{\sqrt{x}}$ ($\gamma_E \approx
0.577215 \dots$ being Euler's constant).
\begin{table*}
  \begin{ruledtabular}
    \begin{tabular}{ccc}
      $(l,m)$ & $\rho_{l m}^{\rm amp\,fit}$ & rel. err. \\[2pt]
      \hline \\[-8pt]
      (2, 2) & $(-20.28 + 12.03\, \textrm{eulerlog}_2 v^2 )\, q v^9$ & $\lesssim 0.3\%$ \\[2pt]
      (2, 1) & $(-0.5144 + 3.175\, \textrm{eulerlog}_1 v^2)\, q^2 v^8$ & $\lesssim 0.4\%$ \\[2pt]
      (3, 3) & $3.894\,q^2 v^8 + (-42.08+ 12.76\, \textrm{eulerlog}_3 v^2 )\, q v^9$ & $\lesssim 0.2\%$ \\[2pt]
      (3, 2) & $-0.6932\, q v^7 -1.558\, q^2 v^8$ & $\lesssim 1\%$ \\[2pt]
      (3, 1) & $-1.012\, q^2 v^8 + (0.8846 -1.279\, \textrm{eulerlog}_1 v^2)\, q v^9$ & $\lesssim 0.08\%$ \\[2pt]
      (4, 4) & $0.9625\, q v^7 + (-2.069 -0.7846\, \textrm{eulerlog}_4 v^2)\, v^8 -0.2633\, q^2 v^8$ & $\lesssim 0.2 \%$ \\[2pt]
      (4, 3) & $1.424\, q^2 v^6 -2.475\, q v^7$ & $\lesssim 0.8\%$ \\[2pt]
      (5, 5) & $(19.51  -5.623\, \textrm{eulerlog}_5 v^2)\, v^6 + 0.3443\, q^2 v^6$ & $\lesssim 1\%$ \\[2pt]
      (6, 6) & $-0.9925\, q v^5 -0.03416\, q^2 v^6 + (19.75 - 5.328\, \textrm{eulerlog}_6 v^2)\, v^6$ & $\lesssim 0.8\%$ \\[2pt]
      (7, 7) & $-1.732\, v^4 + 0.4912\, q^2 v^4 -1.117\, q v^5 + 0.1468\, q^2 v^6 + 
			(25.63 s-6.979\, \textrm{eulerlog}_7 v^2)\, v^6$ & $\lesssim 0.2\%$ \\[2pt]
      (8, 8) & $-0.9946\, q v^3 -0.2949\, v^4 + 
			 0.003748\, q^2 v^4 + 2.428\, q v^5$ & $\lesssim 1.2\%$ \\
     \end{tabular}
  \end{ruledtabular}
  \caption{\label{tab:Fits} Functions $\rho_{l m}^{\rm amp\,fit}$ fitted to individual multipolar modes of the numerical flux at infinity. The coefficients are given with 4 significant figures. In the last column we show the upper bound on the residual relative error of these fits over the spin and frequency ranges used for the fits, i.e. all spins except $q=0.99$, and up to the ISCO.}
\end{table*}
\begin{figure}[!ht]
  \begin{center}
    \includegraphics*[width=0.45\textwidth]{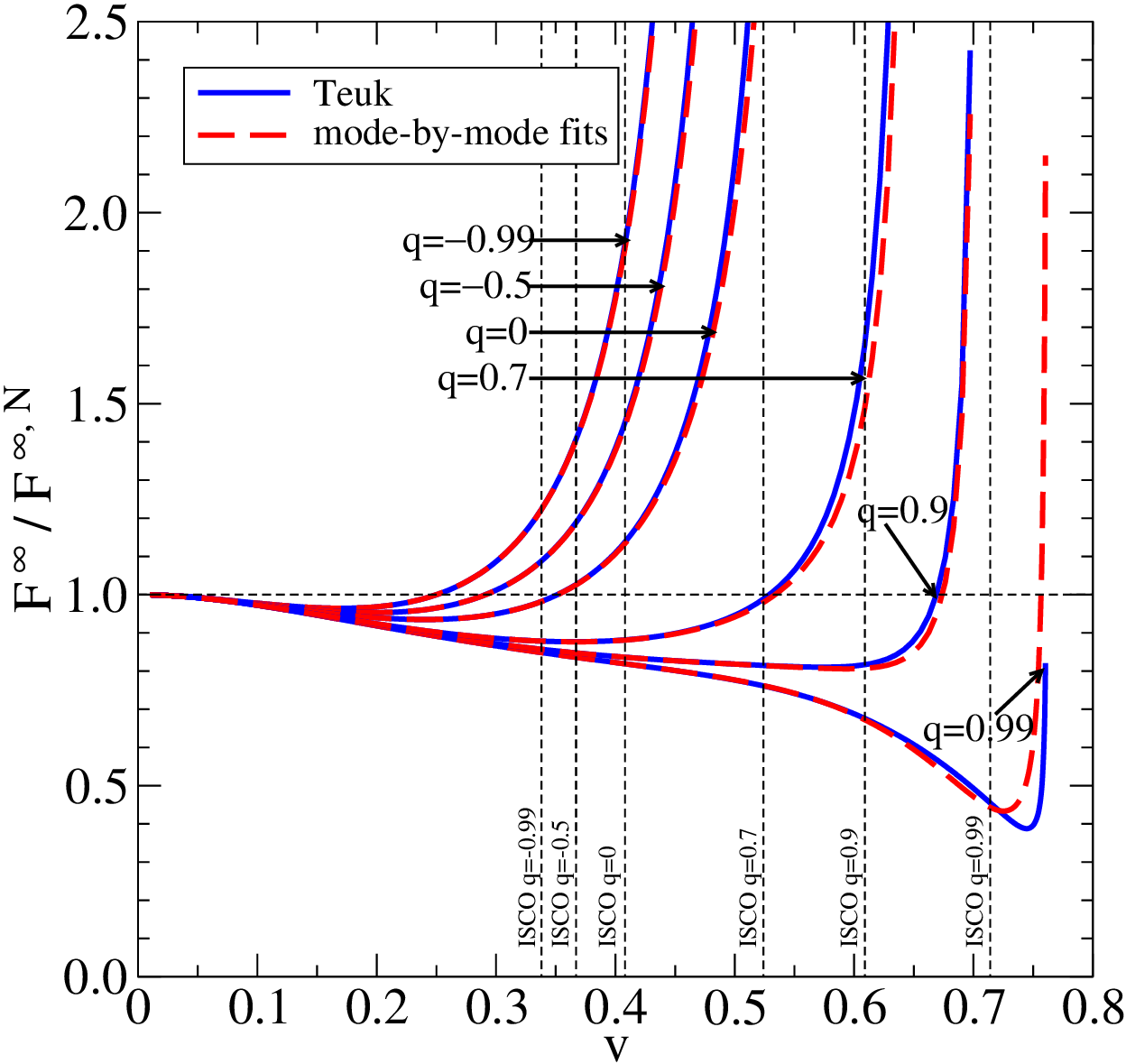}
    \caption{\label{fig:FInfLRFits} We plot the total Teukolsky-equation flux at infinity (in solid blue) and the $\rho_{l m}$--factorized model of Ref.~\cite{Pan2010hz}, improved with the amplitude fits $\rho_{l m}^{\rm amp\,fit}$ (in dashed red). The curves extend up to $r=r_{\rm LR}+0.01M$. The fluxes are normalized by the leading quadrupole luminosity at infinity.}
  \end{center}
\end{figure}

For multipolar modes with $l \leq 4$ the fitting functions contain
only spinning terms. But starting from $l =5 $ both nonspinning and
spinning terms are fitted. For instance, for the (5,5) mode, both the
nonspinning and spinning sector are known through 2.5PN beyond the
leading order, therefore we fit both sectors at 3PN order.

The choice of including logarithmic terms or not is based on the
patterns displayed by the currently available expressions for the
$\rho_{l m}$'s: nonspinning (spinning) logarithmic terms show up at
3PN order beyond the leading nonspinning (spinning) term. We also
choose the spin dependence for the spinning terms to be either linear
of quadratic in $q$, again based on the patterns present in the
$\rho_{l m}$'s: spinning terms proportional to odd (even) powers of
$v$ are odd (even) in the spin $q$.

Finally, the (7,7) and (8,8) modes turn out to be quite difficult to
fit, due to the limited Taylor-expanded knowledge from BH perturbation
theory, and they require as many as 3 PN orders 
to be fitted within a few percent accuracy, which means a total of 6 fitting parameters for
(7,7) and 4 fitting parameters for (8,8). In contrast, all other modes
with $l \leq 6$ can be accurately fitted using only half or one PN
order. We end up fitting a total of 35 coefficients.

The quality of the fits is generally very good on a mode-by-mode
basis, with residuals always smaller than $\sim 1.2\%$ for all the
values of $q$ (except 0.99), for frequencies up to the ISCO and for
all the fitted multipolar modes. In the third column of
Table~\ref{tab:Fits} we list the upper bound for the relative error on
the fits of the multipolar modes.

\begin{figure}[!ht]
  \begin{center}
    \includegraphics*[width=0.45\textwidth]{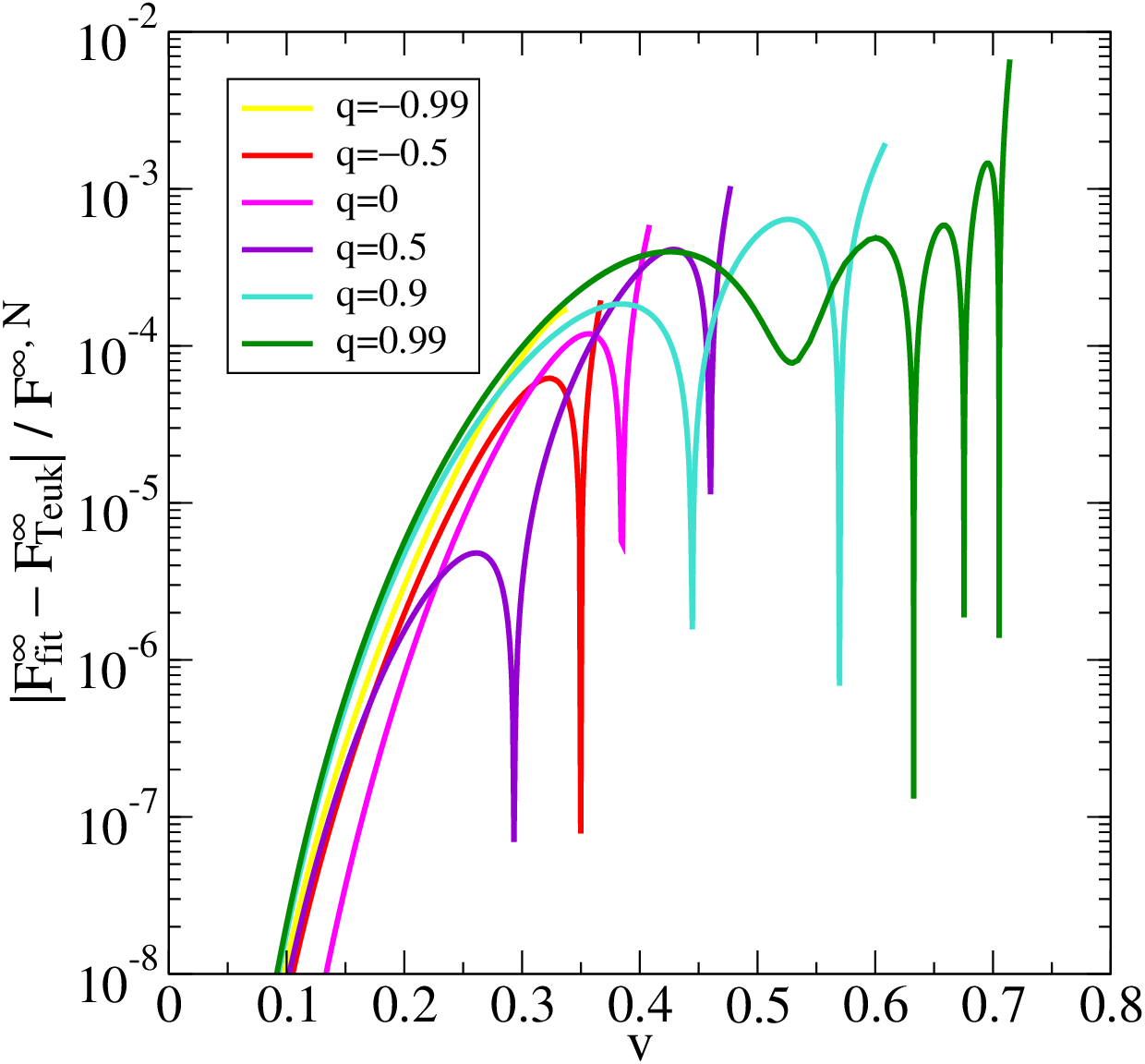}
    \caption{\label{fig:FitFInfRes} We show the absolute residual error (normalized by the leading order luminosity at infinity) on the factorized flux at infinity improved with $\rho_{l m}^{\rm amp\,fit}+\rho_{l m}^{\rm tot\,fit}$. All curves 
      extend up to the respective ISCOs.}
  \end{center}
\end{figure}
We now turn to the total energy flux at infinity. In Fig.~\ref{fig:FInfLRFits} we show comparisons of $F^{\infty}_{\rm Teuk}$ against the model with the mode-by-mode fits discussed above.
When the spins are negative or small, the factorized model of Ref.~\cite{Pan2010hz} actually performs fairly well without any additional fit: for those cases, in fact, the modeling error is less that 1\% at the ISCO, as demonstrated by Fig.~\ref{fig:FInfLR}. In general, the energy flux diverges at the LR since the energy-momentum tensor of the particle sourcing the GW perturbations diverges there as well. This feature is incorporated in the model through the effective source factor $\hat{S}^{(\epsilon)}_{\rm eff}$, which behaves like $(r-r_{\rm LR})^{-1}$ for $r\sim r_{\rm LR}$~\cite{Damour2007,DIN,Pan2010hz}. But, when the spin is large and positive, the divergence of the numerical flux is localized in a narrow neighborhood of the LR, while the model without fits starts growing to large values even before the ISCO. For instance, when $q\geq0.9$, the factorized model differs from the numerical data by more than 100\% even before the ISCO, so that an EOB evolution based on such flux would be unreliable already in the late inspiral, as already pointed out earlier. When the fits are included, the model agrees with the numerical data to within 0.1\% before the ISCO for all the spins up to $q=0.99$, as shown in Fig.~\ref{fig:FInfLRFits}.

As a final refinement, on top of the mode-by-mode fits just discussed, we add 8 additional fitting parameters (4 in the (2,2) mode, 4 in the (3,3) mode), and determine them through a global fit on $F^{\infty}$ itself, similarly to what Refs.~\cite{Gair:2005ih,Yunes:2010zj} did. Again we restrict to $0.01\leq v\leq v_{\rm ISCO}$, but now we include also $q=0.99$. We can achieve a reduction of the error by about an order of magnitude at the ISCO for all the available spins, as shown in Fig.~\ref{fig:FitFInfRes}. These additional terms to be added to $\rho_{l m}+\rho_{l m}^{\rm amp\,fit}$, which we will call $\rho_{l m}^{\rm tot\,fit}$, read
\begin{eqnarray}
\rho_{22}^{\rm tot\,fit} &=& (-9.890+ 9.039\,\textrm{eulerlog}_2 v^2 )\,q^{2}v^{10} \notag\\
&+&
 (-18.84 + 2.486\,\textrm{eulerlog}_2 v^2)\,q v^{11}\,,\\
 \rho_{33}^{\rm tot\,fit} &=&   [73.73 - 
   36.97\,\textrm{eulerlog}_3 v^2 \notag\\
   &+&q^{2}\,(3.955 - 0.7106\,\textrm{eulerlog}_3 v^2 )]\,v^{10}\,.
\end{eqnarray}

\section{Fits of the black-hole absorption gravitational flux}
\label{sect:AppendixFitFH}
\begin{figure}[!ht]
  \begin{center}
    \includegraphics*[width=0.45\textwidth]{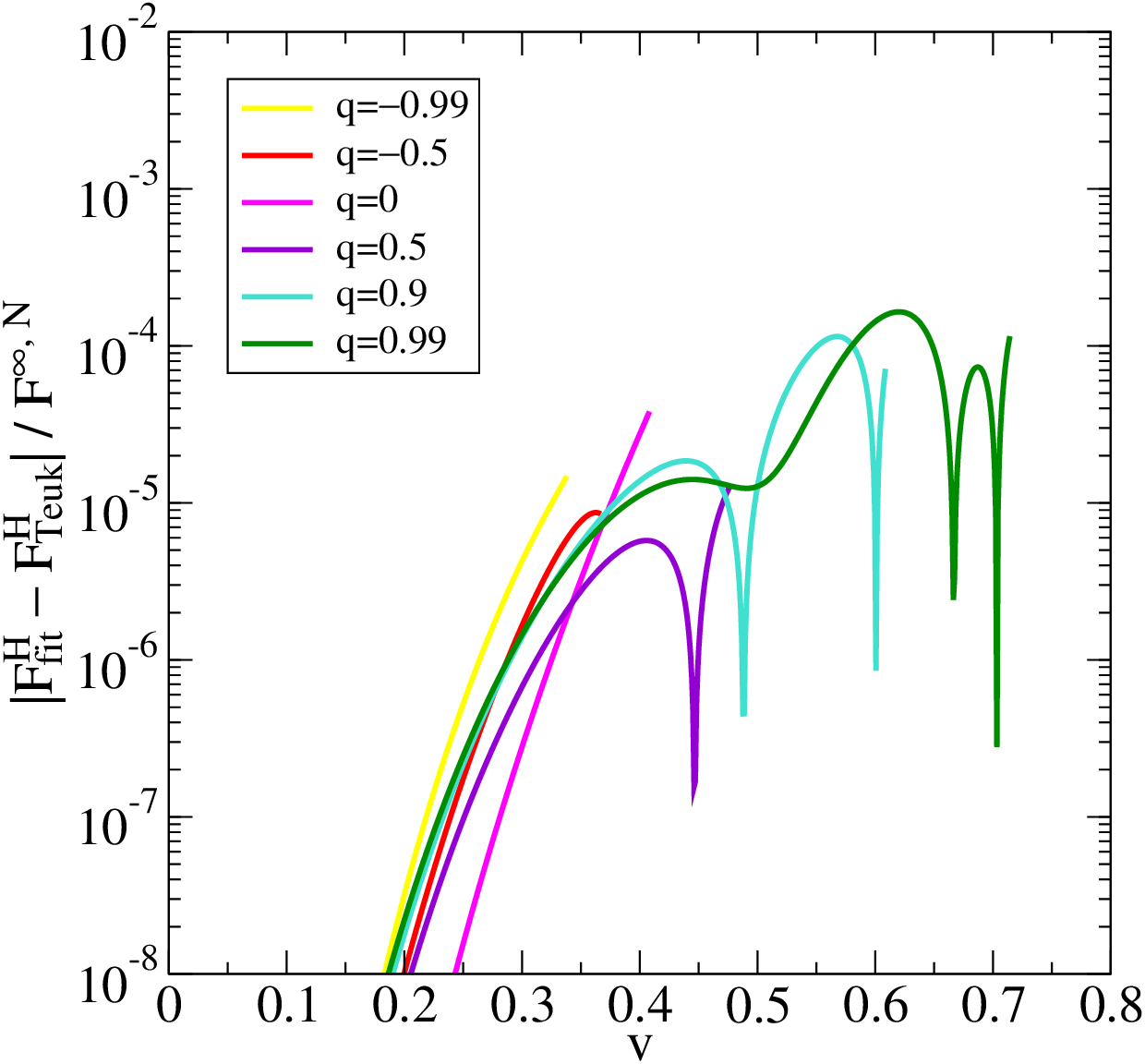}
    \caption{\label{fig:FitFHRes} We show the absolute residual error on the fitted absorption flux, normalized by the leading order luminosity at infinity. All curves 
      extend up to the respective ISCOs.}
  \end{center}
\end{figure}
In this appendix we provide numerical fits to the Teukolsky-equation black-hole absorption fluxes. Our starting point is the $\tilde{\rho}^{\rm H}_{\ell m}$--factorized model developed in this paper. We add to the $\tilde{\rho}^{\rm H}_{\ell m}$'s in Eqs.~(\ref{rho2})--(\ref{rho4}) higher-order PN terms $\tilde{\rho}^{\rm H,fit}_{\ell m}$. In particular, we modify only the dominant and leading subdominant modes (2,2), (2,1) and (3,3). We choose the functional form of the $\tilde{\rho}^{\rm H,fit}_{\ell m}$'s based on the lower PN orders, trying to include similar dependences on $v$ and $q$. We have data for the Teukolsky-equation $F^{\rm H}$ for as many as 22 spins: $q\in\{$$-0.99$, $-0.9$, $-0.8$, $-0.7$, $-0.6$, $-0.5$, $-0.4$, $-0.3$, $-0.2$, $-0.1$, 0, 0.1, 0.2, 0.3, 0.4, 0.5, 0.6, 0.7, 0.8, 0.9, 0.95, 0.99$\}$. The fits are done globally on all spins in $v$--space. The sampled frequency ranges extend from $v=0.01$ up to $r=r_{\rm LR}+0.01M$, but we use data only up $r=(r_{\rm ISCO}+r_{\rm LR}+0.01M)/2$, since attempts to include the whole available velocity ranges spoil the lower frequency portion of the fits; nonetheless our fits prove very accurate up to the ISCO. In order to have residual relative errors within a few percent for all the available spins up to the ISCO, we have to use 11 fitting coefficients. We find
\begin{eqnarray}
\tilde{\rho}^{\rm H,fit}_{22}&=&-(1570+118.5\,
   q+589.7\,\log v)\,v^9\notag\\
   &+&(1323+336.3\,q-1291\,\log
   v)\,v^{10}\,,\\
\tilde{\rho}^{\rm H,fit}_{21}&=&(50.25-54.95\,q-40.39\,\log v)\,v^7\,,\\
\tilde{\rho}^{\rm H,fit}_{33}&=&(15.65-13.41\,q)\,v^{5}\,.
\end{eqnarray}
Figure~\ref{fig:FitFHRes} shows what are the residuals on the fitted
ingoing fluxes, normalized by the leading order luminosity at
infinity. We plot this quantity, rather than the relative residual
errors, because in any realistic setting these fits are going to be
added into a radiation reaction term where the flux at infinity is
also present. In fact, as discussed before (see Fig.~\ref{fig:FHoverFInf}),
$|F^{\rm H}|$ is always much smaller than $|F^{\infty}|$ before the
ISCO, and one is typically interested in an accurate total flux
($F^{\infty}+F^{\rm H}$), hence our choice of the normalization. It is
therefore possible to estimate the modeling error on the total flux by
directly adding Fig.~\ref{fig:FitFInfRes} and Fig.~\ref{fig:FitFHRes}.

\bibliography{References/References}
\end{document}